\documentclass[preprintnumbers,superscriptaddress,nofootinbib,aps,prd,floatfix]{revtex4}

\usepackage{hyperref}
\usepackage{amsmath,slashed}
\usepackage{graphicx}
\usepackage{subfigure}

\newcommand{\be}{\begin{eqnarray*}}
\newcommand{\ee}{\end{eqnarray*}}

\newcommand{\bee}{\begin{eqnarray}}
\newcommand{\eee}{\end{eqnarray}}
\newcommand{\beeq}{\begin{equation}}
\newcommand{\eeeq}{\end{equation}}

\newcommand{\pythia}{\textsc{Pythia}\;}

\DeclareGraphicsExtensions{.pdf,.png,.jpg}

\usepackage{color}

\usepackage{cleveref}
\usepackage{dsfont}
\newcommand{\B}{\mathds{B}}

\begin{document}

\title{Scale and Scheme Variations in Unitarized NLO Merging}

\begin{abstract}
Precision background predictions with well-defined uncertainty estimates are 
important for interpreting collider-physics measurements and for planning 
future high-energy collider experiments.
It is especially important to estimate the perturbative uncertainties in
predictions of inclusive measurements of jet observables, that are 
designed to be largely insensitive to non-perturbative
effects such as the structure of beam-remnants, multi-parton scattering or 
hadronization.
In this study, we discuss possible pit-falls in defining the perturbative uncertainty
of unitarized next-to-leading order multi-jet merged predictions, using
the \pythia event generator as our vehicle. For this purpose, we consider 
different choices of unitarized NLO merging schemes as well as consistent 
variations of renormalization scales in different parts of the calculation.
Such a combined discussion allows to rank the contribution of scale variations
to the error budget in comparison to other contributions due to algorithmic 
choices that are often assumed fixed.
The scale uncertainty bands of different merging schemes largely overlap, but
differences between the ``central" predictions in different schemes can 
remain comparable to scale uncertainties even for very well-separated
jets, or be larger than scale uncertainties in transition
regions between calculations of different jet multiplicity.
The availability of these variations within \pythia will enable more
systematic studies of perturbative uncertainties in precision background
calculations in the future.
\end{abstract}

\author{Leif Gellersen}
\affiliation{Department of Astronomy and Theoretical Physics,\\Lund University, S-223 62 Lund, Sweden}

\author{Stefan Prestel}
\affiliation{Department of Astronomy and Theoretical Physics,\\Lund University, S-223 62 Lund, Sweden}

\preprint{LU-TP-20-03, MCNET-20-04}
\vspace*{4ex}

\maketitle

\section{Introduction}
\label{sec:intro}

Precision predictions of the final states of high-energy scattering signal or 
background processes are crucial for the continued success of
high-energy collider physics. This includes e.g.\ exploiting the potential of
indirect searches for physics beyond the Standard Model (SM) at the Large
Hadron Collider, or precision SM measurements at future lepton colliders. 
The more detailed signal and background final states can be predicted, the 
larger the set of conceivable measurements. General-Purpose Event
generators (GPEGs)~\cite{Buckley:2011ms} produce a detailed description of final states at
the level of individual particles, and thus provide controlled pseudo-data 
(in the form of simulated scattering events) that can be used to develop new
analysis strategies. At the same time, GPEGs aim to predict moderately
exclusive final states with as high precision as possible, such that
precision SM predictions can be juxtaposed with data to allow setting 
exclusion limits on parameters in Beyond-the-SM theories.

The precision of GPEG simulations is typically difficult to quantify,
since the calculations are based on a mix of perturbative calculations
(to determine the distribution of the highest-energy transfer scattering,
and its radiative cascade) and phenomenological models, which are necessary
to describe the scattering final state in detail (e.g. to incorporate
the remnants of colliding beams after scattering, and to ensure conservation
of momentum, electric, and QCD color charges). Some
measurements are constructed to be as insensitive as possible to 
phenomenological beam-remnant, multi-particle scattering or hadronization
models, such that the uncertainties due to perturbative approximations
dominate the overall error budget. Measurements in this category are
very inclusive measurements of inelastic scattering processes that do not, 
at lowest order, include QCD couplings, or moderately inclusive measurements
constructed with the help of infrared safe observables.

The goal of this work is to discuss, define and assess the contribution of
uncertainties due to perturbative approximations to the error budget
of predictions to multi-jet final states at colliders, using precise
next-to-leading (NLO) multi-jet merged predictions within the \pythia event
generator~\cite{Sjostrand:2006za,Sjostrand:2014zea} as a case study. Similar case studies within individual leading-order or NLO
matched predictions have been considered in~\cite{Bellm:2016rhh,Cormier:2018tog}, while~\cite{Bothmann:2016nao}
focussed on the technical validation of variations within a specific
NLO merging scheme. We extend these
discussions by considering the combination of several NLO calculations 
(a particular scheme of combining NLO calculations will 
define what we call an NLO merging scheme), and the uncertainty due to choices in the
combination procedure, i.e. the \emph{merging scheme}. In particular, we focus on the interplay of 
uncertainties from defining exclusive cross sections at higher orders, and scale variations
within unitarized merging schemes, thus addressing the questions: \emph{What
is the impact of choices that are not constrained
by requirements from retaining shower- and/or fixed-order accuracy on the
prediction and uncertainty of the overall calculation? Do some merging schemes
exhibit spuriously sized scale uncertainties?}

\section{Unitarized NLO-merged calculations}
\label{sec:uncertainties}

Scattering events at high-energy colliders such as the LHC potentially 
contain many well-separated jets of particles. To obtain a sound perturbative 
model in such a situation, several precise fixed-order calculations are
necessary to supplement the parton shower, which, due to its ordering
requirement, cannot reach all regions of phase space. Since the parton shower
also relies on approximate collinear/soft splitting kernels, its model
of hard well-separated jets is typically insufficient. Disregarding
hard regions can affect the overall event generator tune, since they can
e.g.\ significantly alter the description of large particle-multiplicty tails.

Thus, several calculations (that are themselves combinations of fixed-order 
and all-order perturbative components) need to be combined. We will refer to
a combination scheme as \emph{merging scheme}\footnote{A relatively complete 
list of matching- and merging methods employed by event generators
can be found e.g.\ in~\cite{Schonherr:2018rlx}.}. The prerequisite for these
are matrix-element generators that can
readily provide the necessary fixed-order calculations. Fixed-order
calculations with light final-state partons require regularization to avoid
infrared singularities. This regularization can be achieved by removing
all phase-space regions for which the value of a kinematically defined 
\emph{merging scale} falls below a pre-defined value. Events that were thus 
discarded can be recovered by subsequent parton showering. Thus, the
\emph{merging scale} takes on a two-fold meaning: It acts as regularization
of fixed-order calculations, and as a separator between fixed-order-
and parton shower phase space regions. Since the merging scale definition and
value are not unique, this introduces an algorithmic uncertainty into
merging schemes. The aim of the current study will be to investigate the 
interplay between theoretically unavoidable uncertainties due to truncation of 
the perturbative series, and higher-order uncertainties in the definition of 
the merging prescription. This interplay can be obscured by varying other
algorithmic choices such as the merging scale. To avoid inconclusive
statements, we will thus not consider such variations below.

Merging schemes rely on consistency conditions designed to ensure that the precision of none
of their parts is degraded: In the phase-space regions in which fixed-order
calculations are supplemented, the fixed-order expansion of the merged
prediction should recover the original calculation, while throughout the
phase-space available to showering, the all-order results should
recover parton-shower resummation. The interplay between these requirements
is especially delicate in ``transition regions", where several components 
contribute almost equally to the final result, and in ``non-shower regions"
beyond the reach of (ordered) showering, for which no comprehensive
all-order description is known.

A straight-forward consistency constraint can be obtained from the unitary
nature of the parton shower, i.e.\ that
the sum of exclusive $n$-parton cross section and inclusive $n+1$-parton
cross section recovers the inclusive $n$-parton cross section for observables
only sensitive to $n$ partons. We can extend this property also to merged 
calculations, thus arriving at unitarized merging 
schemes~\cite{Rubin:2010xp,Lonnblad:2012ng,Lonnblad:2012ix,Platzer:2012bs,Bellm:2017ktr}.

Unitarized NLO merging schemes enforce consistency between different
calculations by removing the complete impact of newly added 
high-parton-multiplicity configurations from the inclusive prediction by 
explicit subtraction of reduced-multiplicity counter-events,
\begin{equation}
\label{eq:unitarization_schematic}
\begin{aligned}
\langle\mathcal{O}\rangle_n
&= ~
\Big( \textnormal{(inclusive rate for $n$ partons)} [\Phi_n] - \textnormal{(inclusive rate for $n+1$ partons)} [\Phi_{n+1}] \Big) \mathcal{O}_n[\Phi_n]\\
&+~~\,\,\textnormal{(inclusive rate for $n+1$ partons)} [\Phi_{n+1}] \mathcal{O}_{n+1}[\Phi_{n+1}] 
\end{aligned}
\end{equation}
where $\mathcal{O}_n$ denotes a fully differential measurement of all momenta in
the state $\Phi_n]$. In reality, this complete removal is only achieved for very specific 
observables (e.g.\ jet observables defined by using the merging scale 
definition as separation criterion, and the inverse of the parton shower
kinematics as recombination scheme), while residual
higher-multiplicty sensitivity remains in observables very different to
the fixed-order regularization cut. Nevertheless, the same cancellation should in principle 
also apply to the uncertainty on the high-parton-multiplicity configurations. The
latter requires very careful handling of all components of the calculation.

A sensible inclusive prediction should also be complemented with an
accurate description of exclusive cross sections that are sensitive to exactly 
$n$ (and only $n$) jets or partons\footnote{These two requirements often
lead to tensions in the definition of the algorithm. As an example,
inclusive correctness of $n+1$-parton states requires power corrections and
the treatment of 4-momentum conservation when sampling each $n+1$-parton state,
whereas exclusive correctness of $n$-parton states is difficult to formally
achieve if recoil effects are present.
}.
Unitarization introduces higher-order 
components that depend on higher-parton-multiplicities into exclusive cross 
sections, by means of the explicit subtraction, i.e.\ through the second term in 
eq.~\ref{eq:unitarization_schematic}. These differ from the naive parton-shower
result by subleading terms in the parton-shower evolution variable. At NLO, 
similar subleading terms can appear by introducing all-order corrections
to (hard) virtual diagrams, i.e.\ through the first term in 
eq.~\ref{eq:unitarization_schematic}. The interplay of these
terms is beyond both NLO fixed-order accuracy as well as shower accuracy.
However, it is of the same order as variations that are used to gauge 
NLO fixed-order uncertainties. 

It is prudent to require a merged calculation
to recover the fixed-order result \emph{as well as the uncertainty} of
the latter in certain regions of phase-space. However, it is not a priori
clear how these regions should be defined, nor that it is obvious that
any merging scheme (taken here to be defined by different choices of reweighting the NLO
corrections) fulfills this requirement. The aim of the current
pilot study is to initiate the discussion of these points. To be
able to discuss subtle changes in the NLO merged event generator predictions
by using a (large) fixed set of statistically produced events, we focus on
``reweightable variations" here, such that the impact of purely statistical 
fluctuations can be minimized. Possible reweightable perturbative variations are 
\begin{itemize}
\item[$a)$] Variations of the renormalization scale, correlated between fixed-order and
parton-shower components;
\item[$b)$] Variations of the (all-order) reweighting of higher-order fixed-order terms.
\end{itemize}
Many other variations are of course possible in an NLO merged calculation. 
However, these might not be reweightable\footnote{\dots as would be the case for factorization scale variations, since 
initial-state parton-shower evolution links factorization scales to the 
phase-space boundaries of the shower, or variation of the event generator tune, 
since different tunes may disallow different perturbative states due to
changes in the shower cut-off.}
and thus require prohibitively large event samples to minimize statistical
effects, or do not have a well-defined perturbative expansion\footnote{\dots
as would be the case for reweightable variations of the PDF set or PDF member,
or non-reweightable merging scale variations.}. Hence, the current study is 
limited to a consistent definition and assessment of the 
variations $a)$ and $b)$ above. Other variations (such as factorization
scale changes) will not invalidate the findings below, and instead might 
serve to put more stringent constraints on the allowed form of NLO merging
schemes.


\section{Theory and Implementation}
\label{sec:implementation}

In this section, we define several variants of unitarized NLO merging strategies 
that have well-motivated, yet different, higher-order structure. For this,
we will start from a very general form of a unitarized merging prescription,
of which the UNLOPS prescriptions~\cite{Lonnblad:2012ix} 
and~\cite{Bellm:2017ktr} are sub-sets. This general starting point allows 
to define several classes of unitarized NLO merging schemes, and thus suggests
a large associated uncertainty. However, most unitarized NLO merging schemes
will have a spurious behavior, e.g.\ if their all-order behavior does not
recover the all-order logarithmic structure of QCD. Thus, we will discuss
conditions that \emph{sensible} new unitarized NLO merging schemes need to 
fulfill, e.g.\ how to define schemes in which scale uncertainties do not deteriorate 
the accuracy of the overall prediction. The result of this is that we 
allow a new source of uncertainty -- the scheme uncertainty -- and determine
sensible scheme variations that may constitute a reasonable assessment of this 
uncertainty.

To begin, it is useful to examine the construction
of \emph{exclusive} jet rates in the merged calculation, with the exclusive 1-jet
rate being a sufficiently complicated example. In a general unitarized calculation,
this rate is given by
\begin{equation}
\label{eq:1jetrate_generic}
\begin{aligned}
\langle\mathcal{O}_1\rangle = 
       \mathcal{O}_1[\Phi_1] &\Bigg( \left(\B_1[\Phi_1] + \B_1^\text{NLO}[\Phi_1]\right)w_\text{NLO}[\Phi_0,\Phi_1] + \B_1[\Phi_1]\left( w_\text{LO}[\Phi_0,\Phi_1] - w_\text{I}[\Phi_0] - w_\text{S}[\Phi_0,\Phi_1] \alpha_s \left\{ w_\text{LO}[\Phi_0,\Phi_1] \right\}_1 \right)\\
       &- \int_{t_\mathrm{ms}} \B_2 [\Phi_2] w_\text{LO}[\Phi_0',\Phi_1'] w_\text{LO} [\Phi_1,\Phi_2] - S_3 \Bigg)
\end{aligned}
\end{equation}
where the factors $w_\text{X}$ are a-priori weights that have to be chosen to
preserve certain accuracy criteria, $\B_i[\Phi_i]$ denote the inclusive
tree-level calculation of the $i$-jet rate, $\B_1^\text{NLO}[\Phi_1]$ all
NLO corrections (including all virtual and real corrections) to the 1-jet
rate, and where we have 
collected pieces stemming from the unitarization of higher-multiplicity contributions 
with lowest order $\alpha_s^3$ into the symbol $S_3$. We assume
that all factors $w_\text{X}$ have a well-defined expansion
\begin{equation}
w_\text{X}(\mu_r) = \sum_{i=0}^\infty \alpha_s^i \left\{ w_\text{X}\right\}_i = 1 + \alpha_s(\mu_r) \left\{ w_\text{X}\right\}_1 + \alpha_s^2(\mu_r) \left\{ w_\text{X}\right\}_2 + \mathcal{O}(\alpha_s^3)
\end{equation}
This expansion immediately guarantees that the lowest-order terms in the
expansion of eq.~\ref{eq:1jetrate_generic} is correctly given by the
tree-level result $\mathcal{O}_1[\Phi_1]\B_1[\Phi_1]$. Similarly, the next term in the
expansion is correctly given by the \emph{exclusive} NLO cross section
\begin{equation}
\langle\mathcal{O}_1\rangle^{(2)} =
\mathcal{O}_1[\Phi_1]\left(\B_1^{NLO}[\Phi_1] - \int_{t_\mathrm{ms}} \B_2 [\Phi_2] \right)
= 
\mathcal{O}_1[\Phi_1]\left( V_1[\Phi_1] + \int^{t_\mathrm{ms}} \B_2 [\Phi_2] \right)
\end{equation}
provided that $\left\{ w_\text{NLO} \right\}_1 = \left\{ w_\text{I}\right\}_1$ and
that the real corrections in $\B_1^\text{NLO}$ and the unitarization term $\int_{t_\mathrm{ms}} \B_2$
are mapped in an identical way to the $\Phi_1$ phase space points. 
Non-unitarized merging schemes arrive at the exclusive NLO cross section 
in a somewhat different manner~\cite{Hoeche:2012yf}.

Note that to achieve the correct behavior for any choice of the
renormalization scale requires $\left\{ w_\text{NLO} \right\}_1 = \left\{ w_{I}\right\}_1$
for each scale value, and thus implies that the expansion of either weight can 
be defined with reference to a common scale, and that
\begin{equation}
  w_\text{X}(\mu_r) - w_\text{X}(k\mu_r) =  \alpha_s^2 (\mu_r) \frac{\beta_0}{2\pi} \ln (k) \left\{ w_\text{X}\right\}_1 + \mathcal{O}(\alpha_s^3)\quad \text{X}\in\{\text{NLO},\text{I}\}\,.
\end{equation}
If this condition is not guaranteed, then the all-order accuracy of the 
prediction, as defined by reference to the parton shower, is compromised.
Changes due to scale variations of weights applied to Born-level contributions 
enter at the same order ($\mathcal{O}(\alpha_s^3)$) as a non-trivial reweighting 
of higher-order (virtual) corrections. 
The interplay between these corrections is the main interest of this article. 
In order to avoid over-generalizations
of conclusions about uncertainties drawn from scale variations, we will analyse
the $\mathcal{O}(\alpha_s^3)$ expansion of 
eq.~\ref{eq:1jetrate_generic} and set up conditions for the reweighting
of higher-order corrections. Any reweighting must not, of course, introduce 
spurious enhacements at any order, since this would exaggerate the
``scheme variation". With this in mind, different reweighting strategies
can be used together with scale variations to determine more robust
uncertainties.

The $\mathcal{O}(\alpha_s^3)$ expansion of 
eq.~\ref{eq:1jetrate_generic} reads~\cite{Lonnblad:2012ix}
\begin{equation}
\label{eq:1jetrate_as3exp}
\begin{aligned}
\langle\mathcal{O}_1\rangle^{(3)} = 
&\mathcal{O}_1[\Phi_1]\Bigg(
\left\{ w_\text{NLO}\right\}_2 \B_1
+
\left\{ w_\text{NLO}\right\}_1 \B_1^\text{NLO}
+
\left\{ w_\text{LO}\right\}_2 \B_1
-
\left\{ w_\text{I}\right\}_2 \B_1
-
\left\{ w_\text{S}\right\}_1 \left\{ w_\text{LO}\right\}_1 \B_1\\
&\quad\qquad-\int_{t_\mathrm{ms}}\Bigg(
 \B_2^\text{NLO}
- \int \B_3^{\downarrow} \Bigg)
\Bigg)~,
\end{aligned}
\end{equation}
where we have explicitly included the relevant $S_3$ terms and assumed
that $\left\{ w_\text{NLO} \right\}_1 = \left\{ w_\text{I}\right\}_1$ also applies 
to the reweighting of two-parton corrections. 

It is reasonable to expect that eq.~\ref{eq:1jetrate_as3exp} reproduces
the correct coefficient of the largest contribution for
the observable $\mathcal{O}_1[\Phi_1]$. If the observable measures the
merging scale value of states in $\Phi_1$, we expect a behavior $
\mathcal{O}_1^{(3)} \propto \alpha_s^3\ln^6(t_\mathrm{ms})$. Since 
$\left\{ w_\text{LO}\right\}_2 \B_1 \propto \alpha_s^3\ln^6(t_\mathrm{ms})$, this amounts
to constraints on and/or cancellations between the different weights in
eq.~\ref{eq:1jetrate_as3exp}. The most straight-forward way to
enforce such cancellations is to set 
$w_\text{NLO} = w_\text{I}$ for any scale value (we have already seen that this holds for the lowest
and next-to-lowest order), and then constrain the concrete form of $w_\text{NLO}$
from the remaining terms. If we assume that $ \left\{ w_\text{LO}\right\}_1 \B_1$ 
provides a sensible approximation of the leading parts of $\B_1^\text{NLO}$, it
is tempting to identify
\begin{equation}
\label{eq:b1nlo_rwgt}
\left\{ w_\text{NLO}\right\}_1 \B_1^\text{NLO} 
-
\left\{ w_\text{S}\right\}_1 \left\{ w_\text{LO}\right\}_1 \B_1
\longrightarrow
\left\{ w\right\}_1 \left[\B_1^\text{NLO} - \left\{ w_\text{LO}\right\}_1 \B_1\right]
\quad\textnormal{i.e.}\quad \left\{ w_\text{S}\right\}_1 = \left\{ w_\text{NLO}\right\}_1 = \left\{ w\right\}_1 (=\left\{ w_\text{I}\right\}_1) ~.
\end{equation}
As long as $\left\{ w\right\}_1$ scales as $ \alpha_s\ln^2(t_\mathrm{ms})$ or
less, and while the difference in brackets is subleading, this combination 
will not lead to undesirable enhancements. Nevertheless, while 
$\left\{ w\right\}_1\neq 0$ and $\B_1^\text{NLO} \neq \left\{ w_\text{LO}\right\}_1 \B_1$,
this contribution is a new, process-dependent, source of scale uncertainties 
beyond the NLO and parton shower approximations. The impact of this term 
on the overall uncertainty is thus best estimated by considering explicit 
test cases, as e.g.\ done in the next section.
It could be argued that having $w \not\equiv 1$ brings the merging prescription 
closer to the traditional separation into all-order $W$-terms and 
fixed-order $Y-$terms in analytic resummation~\cite{Collins:1981va,Collins:1984kg}, which typically includes hard 
virtual corrections into the all-order $W$-term~\cite{Collins:1981uk} and assumes that the fixed-order
$Y$-term not only remains constant, but vanishes in the limit when 
$\Phi_1$ and $\Phi_2$ become indistinguishable~\cite{Catani:2010pd}.
It is however not directly obvious that the calculation in~\cite{Collins:1981uk} 
translates to the context of a fully differential event generator employing
IR/UV regularization prescriptions different from~\cite{Collins:1981uk}. 
An assessment of the numerical effect of different treatments is thus
interesting in its own right, also beyond the context of 
its interplay with renormalization scale variations in NLO merging schemes.

To summarize, the above considerations lead to a simple guideline how higher-order corrections
may be reweighted without compromising the quality of the calculation: All terms
in the expansion of the leading-order reweighting $w_\text{LO}$, and all NLO
corrections, should be reweighted by the same (potentially dynamical) weight. Enhancements
appearing in the expansion of this weight should be in agreement with standard 
QCD expectations.
Below, we describe different unitarized NLO merging strategies that meet
these criteria. We then assess the uncertainties on predictions that result 
from consistent renormalization scale variations in matrix element 
generation, merging and parton shower, as well as the ``merging scheme"
uncertainty. We limit the discussion to predictions that are NLO correct up to the first 
additional jet with respect to the reference process, and LO correct for the 
second and third jet. The generalization to higher multiplicity is straight 
forward, but omitted here in favor of readability.

\subsection{UNLOPS}
We start from the UNLOPS multi-leg NLO merging scheme described in 
detail in \cite{Lonnblad:2012ix}. The expectation value of an arbitrary 
jet observable $\mathcal{O}$ in UNLOPS is given by
\begin{equation}
\begin{aligned}
  \mathcal{O}_0 &\left( \bar \B_0 - \int \B_1\left(\Pi_0(k) w_{f,0} \frac{\alpha_s(kp_{\perp,1})}{\alpha_s(k\mu_\mathrm{R})}K-1-\left\{\Pi_0(k)\right\}_{\alpha_s(k\mu_\mathrm{R})} -\{w_{f,0}\}_{\alpha_s(k\mu_\mathrm{R})} - \left\{\frac{\alpha_s(kp_{\perp,1})}{\alpha_s(k\mu_\mathrm{R})} \right\}_{\alpha_s(k\mu_\mathrm{R})} - \{K\}_{\alpha_s(\mu_\mathrm{R})}\right)\right.\\
  &\left.-\int \bar \B_1 \right) \\
  + \mathcal{O}_1 &\left( \bar \B_1 + \B_1\left( \Pi_0(k) w_{f,0} \frac{\alpha_s(kp_{\perp,1})}{\alpha_s(k\mu_\mathrm{R})}K - 1 - \left\{ \Pi_0(k) \right\}_{\alpha_s(k\mu_\mathrm{R})} -\{w_{f,0}\}_{\alpha_s(k\mu_\mathrm{R})}- \left\{\frac{\alpha_s(kp_{\perp,1})}{\alpha_s(k\mu_\mathrm{R})} \right\}_{\alpha_s(k\mu_\mathrm{R})}- \{K\}_{\alpha_s(\mu_\mathrm{R})}\right) \right. \\ 
          &\left. - \int \B_2 \Pi_0(k) w_{f,0} \frac{\alpha_s(kp_{\perp,1})}{\alpha_s(k\mu_\mathrm{R})} \Pi_1(k) w_{f,1} \frac{\alpha_s(kp_{\perp,2})}{\alpha_s(k\mu_\mathrm{R})}K\right) \\
          + \mathcal{O}_2 &\int \B_2 \Pi_0(k) w_{f,0} \frac{\alpha_s(kp_{\perp,1})}{\alpha_s(k\mu_\mathrm{R})} \Pi_1(k) w_{f,1} \frac{\alpha_s(kp_{\perp,2})}{\alpha_s(k\mu_\mathrm{R})}K \, . 
        \end{aligned}\label{eq:unlops}
\end{equation}
with $k=1$. In the above equation, $\B$ and $\bar\B$ denote fully differential cross 
sections at leading and next-to-leading order in the strong
coupling, in the following also called matrix element samples. These are interfaced using the LHEF format. $\{X\}_{\alpha_\mathrm{s}}$ denotes the $\mathcal{O}(\alpha_\mathrm{s})$ contribution to the term $X$. $\Pi_n(k)$ is short for 
\begin{equation}
  \Pi_n(p_{\perp,n},p_{\perp,n+1};x,k) = \prod_j\exp \left\{ -\sum_{i} \int_{p_{\perp,n+1}}^{p_{\perp,n}} \frac{\text d p_\perp^2}{p_\perp^2}\int\frac{\text d z}{z}\frac{\alpha_s(k p_\perp)}{2\pi}P_{ji}(z)\frac{f_i(x/z,p_\perp)}{f_j(x,p_\perp)}\right\}
\end{equation} 
and describes the no-emission probability between the two evolution scales, 
taking into account all allowed $i\rightarrow j$ splittings of all legs in an
$n$ parton state, as described by the kernels $P_{ji}(z)$. This no-emission
probability can be calculated numerically by trial parton showering~\cite{Lonnblad:2001iq}. 
In \pythia 8 \cite{Sjostrand:2014zea}, where the evolution of partons by emissions 
and the simulation of secondary multiparton interactions (MPI) is 
described by a common, interleaved, evolution sequence, the no-emission 
probabilities generated by trial showering can also accommodate no-MPI 
probabilities of the relevant transverse momentum scales~\cite{Lonnblad:2011xx}. 
This allows a smooth combination of input matrix element samples with the
MPI machinery. 

In case of hadronic initial states, the necessary PDF ratios 
\begin{equation}
  w_{f,n} = \frac{x_n^+f_n^+(x_n^+,p_{\perp,n})}{x_n^+f_n^+(x_n^+,\mu_F)}\frac{x_n^-f_n^-(x_n^-,p_{\perp,n})}{x_n^-f_n^-(x_n^-,\mu_F)}\prod_{i=1}^n\frac{x_{i-1}^+f_{i-1}^+(x_{i-1}^+,p_{\perp,{i-1}})}{x_{i-1}^+f_{i-1}^+(x_{i-1}^+,p_{\perp,i})}\frac{x_{i-1}^-f_{i-1}^-(x_{i-1}^-,p_{\perp,{i-1}})}{x_{i-1}^-f_{i-1}^-(x_{i-1}^-,p_{\perp,i})}
\end{equation}
are included as additional weights. As argued
above, we do not vary the factorization scale, since it is not obvious how
to achieve a consistent reweightable variation in this case. The subscript on 
$\mathcal{O}_n$ denotes the final state multiplicity of the states handed to
subsequent event generation steps. Thus, in some cases, e.g.\ to create
counter-events for unitarization, the state used to
evaluate the matrix element calculation is replaced by a lower-multiplicty
state determined by reclustering according to a reconstructed parton shower history.
Finally, the $K$-factor is only applied to configurations that could have been 
produced by parton shower emissions. These regions are defined by being able 
to construct at least one parton shower history of emissions that are ordered
in a decreasing sequence of evolution scales. This choice is consistent with 
the MC@NLO matching scheme, where hard real-emission configurations not reachable
by showering are described with tree-level matrix elements.

Note that in this UNLOPS scheme, we set $w_{S} =  w_{NLO} = w_I \equiv 1$, i.e.\ treat
all higher-order fixed-order corrections as ``hard corrections" that
do not contribute to the all-order result. This relatively conservative approach
has the advantage that scale variations due to ``hard" virtual corrections
do not introduce all-order uncertainties. It has the disadvantage that any ``soft"
virtual correction terms in $\B_1^{NLO}$ that are not correctly reproduced by 
the parton shower are not leveraged to define a more realistic all-order 
uncertainty.

We may estimate the theoretical uncertainties of predictions obtained by the UNLOPS NLO 
merging procedure by variations of the renormalization 
scale $\mu_\mathrm{R}$. For consistency, these variations should not be 
limited to the seed cross sections, but should also include renormalization
scale variations of the parton shower.
For this reason, we employ the scale variations that have been implemented in 
the \pythia 8 parton shower \cite{Mrenna:2016sih}. We extend this
procedure to ensure consistent simultaneous variations in the calculation of 
merging weights, and a consistent set-up between matrix-element and parton-shower
contributions. Every renormalization scale in the matrix element samples 
$\B_n$, $\bar \B_n$ is varied, as is every explicit occurrence 
of $\mu_\mathrm{R}$ in the above formula. Furthermore, the same variation is 
applied to each argument of $p_{\perp,n}$ in the strong coupling 
$\alpha_\mathrm{s}$ in the parton shower. The variation can be produced by 
reweighting using the variation factor $k$ in \cref{eq:unlops}

Note that in this particular NLO merging prescription, the NLO corrections 
$\bar \B_n$ are not reweighted. Below, we will consider variants in which NLO
events are weighted in different a manner.

\subsection{UNLOPS-P}

Alternative unitarized merging schemes that remain NLO correct and do not
degrade the all-order behavior can be obtained by suitably changing how
all-order weights are applied to higher-order fixed-order contributions. In
eqs.~\ref{eq:1jetrate_as3exp}-\ref{eq:b1nlo_rwgt}, we have argued that
choosing a common reweighting for the NLO corrections $\B_1^{NLO}$ and
the $\mathcal{O}(\alpha_s)$ expansion of the parton shower is one (simple)
way to comply with all accuracy constraints.

As first alternative to UNLOPS, we will define the UNLOPS-P scheme (where the
``P" is intended to signify the extended use of no-emission 
\emph{p}robabilities). This alternative unitary merging scheme is inspired 
by the treatment of higher-multiplicity NLO corrections in the UN${}^2$LOPS NNLO 
matching prescription~\cite{Hoeche:2014aia}, and amounts to applying a Sudakov 
weight factor (consisting of PDF ratios and  no-emission probabilities) to the 
higher order terms. In~\cite{Hoeche:2014aia}, it was argued that reweighting the
remnant bracket in eq.~\ref{eq:b1nlo_rwgt} with a Sudakov factor can be 
interpreted as dressing an IR-subtracted hard state with the effects of
soft and collinear radiation. In UNLOPS, the IR-subtracted NLO correction
is instead not dressed with higher-order effects. 
The use of Sudakov factors could be regarded more physical. It has
the added benefit that the impact NLO corrections to 1-jet states in soft/collinear
regions is reduced, thus leading to a gain in numerical stability for small 
merging scale values. 
In the UNLOPS-PC scheme below, we will reassess and refine the interpretation as ``dressing with the effects of radiation".

The expectation value of an arbitrary 
jet observable $\mathcal{O}$ in UNLOPS-P is given by
\begin{equation}
\begin{aligned}
\mathcal{O}_0 &\left( \bar \B_0 - \int \B_1\Pi_0(k) w_{f,0} \left(\frac{\alpha_s(kp_{\perp,1})}{\alpha_s(k\mu_\mathrm{R})}K-1-\left\{\Pi_0(k)\right\}_{\alpha_s(k\mu_\mathrm{R})} -\{w_{f,0}\}_{\alpha_s(k\mu_\mathrm{R})} - \left\{\frac{\alpha_s(kp_{\perp,1})}{\alpha_s(k\mu_\mathrm{R})} \right\}_{\alpha_s(k\mu_\mathrm{R})} \right. \right. \\
&\left.\left. - \{K\}_{\alpha_s(\mu_\mathrm{R})} \right.\Bigg) -\int \bar \B_1\Pi_0(k) w_{f,0} \right) \\
      + \mathcal{O}_1 &\left( \bar \B_1 \Pi_0(k)w_{f,0}+ \B_1 \Pi_0(k)w_{f,0} \left(  \frac{\alpha_s(kp_{\perp,1})}{\alpha_s(k\mu_\mathrm{R})}K - 1 - \left\{ \Pi_0(k) \right\}_{\alpha_s(k\mu_\mathrm{R})} -\{w_{f,0}\}_{\alpha_s(k\mu_\mathrm{R})} - \left\{\frac{\alpha_s(kp_{\perp,1})}{\alpha_s(k\mu_\mathrm{R})} \right\}_{\alpha_s(k\mu_\mathrm{R})} \right.\right. \\
    & \left.\left.-\{K\}_{\alpha_s(\mu_\mathrm{R})}\right.\Bigg) - \int \B_2 \Pi_0(k) w_{f,0}\frac{\alpha_s(kp_{\perp,1})}{\alpha_s(k\mu_\mathrm{R})} \Pi_1(k) w_{f,1} \frac{\alpha_s(kp_{\perp,2})}{\alpha_s(k\mu_\mathrm{R})}K\right) \\
          + \mathcal{O}_2 &\int \B_2 \Pi_0(k) w_{f,0} \frac{\alpha_s(kp_{\perp,1})}{\alpha_s(k\mu_\mathrm{R})} \Pi_1(k) w_{f,1} \frac{\alpha_s(kp_{\perp,2})}{\alpha_s(k\mu_\mathrm{R})} K
\end{aligned}
\end{equation}
The double logarithmic Sudakov factor is dominant in the soft/collinear region,
suppressing the rise of the $\mathcal{O}(\alpha_s)$ corrections. These
features should be noticeable both in $1$-jet inclusive observables, but 
also, by virtue of unitarization, in exclusive $0$-jet observables.

\subsection{UNLOPS-PC}

Another alternative to UNLOPS is the UNLOPS-PC scheme defined in the following 
(where the ``C" is intended to signify the extended use of running-\emph{c}oupling
factors). This scheme is motivated by clarifying the 
argument of~\cite{Hoeche:2014aia}: that reweighting the
remnant bracket in eq.~\ref{eq:b1nlo_rwgt} can be 
interpreted as dressing a IR-subtracted hard state with the effects of
soft and collinear radiation. In UN${}^2$LOPS and UNLOPS-P, it was assumed that
the latter effects can be approximated through the application of Sudakov 
factors. Sudakov factors primarily encapsulate the dressing
of parton propagators with self-energy corrections. However, a systematic
treatment of leading-logarithmic dressing also includes the effect of vertex 
corrections, and of running-coupling effects~\cite{Amati:1978by} to obtain
an approximation of the correct ladder diagrams. It thus stands to reason that
a more physical notion of ``dressing with the effects of radiation" should include
both Sudakov- and running-coupling reweighting. This constitutes the 
UNLOPS-PC scheme, in which the expectation value of an arbitrary 
jet observable $\mathcal{O}$ is given by
\begin{equation}
\begin{aligned}
\mathcal{O}_0 &\left( \bar \B_0 - \int \B_1 \Pi_0(k) w_{f,0} \frac{\alpha_s(kp_{\perp,1})}{\alpha_s(k\mu_\mathrm{R})} \left( K-1-\left\{\Pi_0(k)\right\}_{\alpha_s(k\mu_\mathrm{R})} -\{w_{f,0}\}_{\alpha_s(k\mu_\mathrm{R})} - \left\{\frac{\alpha_s(kp_{\perp,1})}{\alpha_s(k\mu_\mathrm{R})} \right\}_{\alpha_s(k\mu_\mathrm{R})} \right.\right. \\
&\left.\left.- \{K\}_{\alpha_s(\mu_\mathrm{R})} \right.\Bigg) -\int \bar \B_1\Pi_0(k) w_{f,0} \frac{\alpha_s(kp_{\perp,1})}{\alpha_s(k\mu_\mathrm{R})} \right) \\
+ \mathcal{O}_1 &\left.\Bigg( \bar \B_1\Pi_0(k) w_{f,0} \frac{\alpha_s(kp_{\perp,1})}{\alpha_s(k\mu_\mathrm{R})} + \B_1 \Pi_0(k) w_{f,0} \frac{\alpha_s(kp_{\perp,1})}{\alpha_s(k\mu_\mathrm{R})}\left.\Bigg(K - 1 - \left\{ \Pi_0(k) \right\}_{\alpha_s(k\mu_\mathrm{R})} -\{w_{f,0}\}_{\alpha_s(k\mu_\mathrm{R})} \right.\right. \\
    &\left.\left. - \left\{\frac{\alpha_s(kp_{\perp,1})}{\alpha_s(k\mu_\mathrm{R})} \right\}_{\alpha_s(k\mu_\mathrm{R})}  -\{K\}_{\alpha_s(\mu_\mathrm{R})}\right.\Bigg) - \int \B_2 \Pi_0(k) w_{f,0} \frac{\alpha_s(kp_{\perp,1})}{\alpha_s(k\mu_\mathrm{R})} \Pi_1(k) w_{f,1} \frac{\alpha_s(kp_{\perp,2})}{\alpha_s(k\mu_\mathrm{R})}K\right) \\
          + \mathcal{O}_2 &\int \B_2 \Pi_0(k) w_{f,0} \frac{\alpha_s(kp_{\perp,1})}{\alpha_s(k\mu_\mathrm{R})} \Pi_1(k) w_{f,1} \frac{\alpha_s(kp_{\perp,2})}{\alpha_s(k\mu_\mathrm{R})}K
\end{aligned}
\end{equation}
This scheme treats leading order and next-to-leading order contributions on 
equal footing. The inclusion of no-emission probabilities regulates the 
contribution of radiative events in soft/collinear regions of phase space. The 
inclusion of the strong coupling ratio produces an opposing effect, increasing
the impact of NLO corrections at lower splitting scales. Due to the exponential 
form of the no-emission probability, the Sudakov suppression will naturally 
overcome the coupling ratio effect at lower scales. Nevertheless, away from 
the collinear limit, the single logarithmic evolution of the strong coupling
ratio is not negligible compared to the Sudakov double logarithm.

\subsection{Comparison of +1j contributions}

Before continuing, it is useful to reiterate the differences between the UNLOPS 
variants, in order to gain some intuition about the impact on phenomenology.
The differences mostly pertain to the treatment of the $+1$ jet 
contribution.
For clarity, we split the Born+virtual+real contribution $\bar \B_i$ into its 
LO component $\B_i$ and a pure NLO correction $\B_i^\text{NLO}$, and label the 
original UNLOPS prescription as UNLOPS-1, since a unit weight is applied to NLO 
contributions. With the notation
\begin{equation}
  w_1 = \Pi_0(k) w_{f,0} \frac{\alpha_s(kp_{\perp,1})}{\alpha_s(k\mu_\mathrm{R})}K\, ,
\end{equation}
the $+1$ jet components of the merging schemes read
\begin{description}
  \item[UNLOPS-1]
    \begin{equation}
      \B_1 w_1 + \left[ \B_1^\text{NLO} - \B_1\left(\left.\Pi_0(k)\right|_{\alpha_s(k\mu_\mathrm{R})} +w_{f,0}|_{\alpha_s(k\mu_\mathrm{R})} + \left.\frac{\alpha_s(kp_{\perp,1})}{\alpha_s(k\mu_\mathrm{R})} \right|_{\alpha_s(k\mu_\mathrm{R})}+ K|_{\alpha_s(\mu_\mathrm{R})}\right) \right]\label{eq:unlops1}
        \end{equation}
  \item[UNLOPS-P]
    \begin{equation}
      \B_1 w_1 + \left[ \B_1^\text{NLO} - \B_1\left(\left.\Pi_0(k)\right|_{\alpha_s(k\mu_\mathrm{R})} +w_{f,0}|_{\alpha_s(k\mu_\mathrm{R})} + \left.\frac{\alpha_s(kp_{\perp,1})}{\alpha_s(k\mu_\mathrm{R})} \right|_{\alpha_s(k\mu_\mathrm{R})}+ K|_{\alpha_s(\mu_\mathrm{R})}\right) \right] \Pi_0(k) w_{f,0}\label{eq:unlopsp}
    \end{equation}
  \item[UNLOPS-PC]
    \begin{equation}
        \B_1 w_1 + \left[ \B_1^\text{NLO} - \B_1\left(\left.\Pi_0(k)\right|_{\alpha_s(k\mu_\mathrm{R})} +w_{f,0}|_{\alpha_s(k\mu_\mathrm{R})} + \left.\frac{\alpha_s(kp_{\perp,1})}{\alpha_s(k\mu_\mathrm{R})} \right|_{\alpha_s(k\mu_\mathrm{R})}+ K|_{\alpha_s(\mu_\mathrm{R})}\right) \right] \Pi_0(k) w_{f,0} \frac{\alpha_s(kp_{\perp,1})}{\alpha_s(k\mu_\mathrm{R})} \label{eq:unlopspc}
    \end{equation}
\end{description}

Thus, the main difference between the variants lies in the factor multiplying the term 
in square brackets. As argued above, it is important to apply weights to this
\emph{combined} term, since, if a logarithmically enhanced weight is applied 
to only the NLO term, or only the product of Born-term and the first-order 
expanded weight, then a leading-logarithmic term will be introduced on higher orders in $\alpha_s$, 
thus spoiling the LL accuracy of the merging prescription. 

For well-separated hard emissions, the UNLOPS-1 and UNLOPS-P schemes should agree, since the 
Sudakov factor is approaching unity. This does not necessarily extend to the 
UNLOPS-PC scheme, for which $\alpha_s(p_{\perp,1}) / \alpha_s(\mu_\mathrm{R}) \neq 1$ 
is possible if $p_\perp\not\rightarrow\mu_\mathrm{R}$ for increasingly hard 
emissions. This is e.g. the case in $e^+ e^-\rightarrow$ jets processes in
\textsc{Pythia}, where the emission $p_\perp$ in final state radiation is bounded by 
$p_\perp < m_{e^+ e^-}/2$, and $\mu_\mathrm{R}$ is typically set to 
$m_{e^+ e^-}$, such that the ratio is strictly larger than one. 
The ratio may also be smaller than one if parton shower emissions or 
reconstructed histories are possible at higher $p_\perp$ values than the 
$\mu_\mathrm{R}$, as is e.g. the case in Drell-Yan events with a jet 
$p_\perp > \mu_\mathrm{R} = M_\mathrm{Z}$.

The term in brackets can, depending on kinematics, have either sign. Thus, it
is not immediately obvious if changing from one UNLOPS variant to another
will uniquely lead to either enhancement or depletion. However, we expect the 
UNLOPS-P prediction to be closest to a leading-order (UMEPS) result, since 
the Sudakov weight is smaller than unity.
Due to the application of the Sudakov factor in UNLOPS-P, the fraction of negative 
cross section in the event generation is expected to be reduced. In UNLOPS-PC, 
the Sudakov factor and the coupling ratio weight have opposite effects on the
fraction of negative cross section, so that the net effect is not obvious.
For the specific example of $e^+e^- \rightarrow \text{jets}$ at the center of 
mass energy $\sqrt{s} = M_\mathrm{Z}$, with up to one additional jet at NLO, 
and a merging cut at $\min_{ij}p_{\perp,ij}^\mathrm{Lund} = 5\,\text{GeV}$, we find
that the fractions of negative contributions to the cross section are given by
\begin{center}
\begin{tabular}{c|c|c|c}
  & UNLOPS-1 & UNLOPS-P & UNLOPS-PC \\ \hline
  $\frac{|\sigma_\mathrm{incl}^-|}{|\sigma_\mathrm{incl}^-| +|\sigma_\mathrm{incl}^+|}$ & 35.7\% & 32.4\% & 36.2\% 
\end{tabular}\, .
\end{center}
Thus, the amount of negative contributions differs only very mildly between
the schemes.

\section{Application and Results}
\label{sec:results}

This section intends to assess the impact of renormalization scale 
and merging scheme variations using a small selection of illustrative example observables.
We have implemented the different variants of unitary NLO merging in \pythia 8, 
relying on matrix element input from MadGraph5\_aMC@NLO \cite{Alwall:2014hca}.
Furthermore, we implemented the renormalization scale
variation, taking into account variations of fixed-order and
parton shower origin, as well as in the weights applied in the merging
procedure. The implementation will be made available in a future release of 
\pythia 8.3.

For reasons of consistency between the parton-shower subtraction terms employed 
in aMC@NLO and the event generator, we use a non-default configuration of the 
parton shower for the first emission. This includes a global recoil scheme, 
where the recoil of a final-state emission is shared among all final state 
partons in the event. Furthermore, matrix element corrections to the parton
shower are removed, and we do not allow for an $\alpha_\mathrm{s}$ running in 
the first parton shower emission, since we use fixed renormalization scale 
choices in aMC@NLO when generating matrix elements. 
We terminate the evolution after the first emission, and store the resulting
events as Les-Houches event files. Using these settings ensures
that the parton shower contribution of the first emission on Born 
configurations correctly cancels the parton-shower subtraction terms used in 
the generation of matrix elements, leading to a consistent NLO event sample.
This event sample is then used as input for subsequent NLO merging, which
proceeds using a default \pythia shower setup.
For consistency with the scale variations performed in the matrix element 
generation, which are based on the $\alpha_\mathrm{s}$ running provided by 
the employed PDF packages, we use a second order running coupling with 
reference value of $\alpha_\mathrm{s}(M_\mathrm{Z}) = 0.118$. This allows us 
to consistently vary renormalization scales  
within matrix element generation, merging weight evaluation and 
parton shower emissions.

To generate events, we employ the minimal value of all possible shower 
splitting scales $p_{\perp,ij}^\mathrm{Lund}$ as a merging scale\footnote{See
e.g.~\cite{Lonnblad:2012ng} for a detailed definition.} to regularize fixed-order inputs and act as separator between the hard emission 
phase space described by the fixed order matrix element, and the soft region 
described by the parton shower. 

To assess uncertainties, we investigate the effect of variations on the
modelling of the $e^+ e^- \rightarrow \;\text{jets}$ and 
$p p \rightarrow W + \text{jets}$ processes. For both processes, we include
up to one additional jet at NLO and the second and third jet emission at LO 
fixed order accuracy. The plots in this section are generated using the \textsc{Rivet} 
toolkit \cite{Buckley:2010ar}. In order to suppress large fluctuations in the 
variation bands at low scales, we do not allow for shower variations below 
three times the shower cut off, and limit the allowed range of variation of the 
strong coupling by requiring $|\alpha_\mathrm{s}'-\alpha_\mathrm{s}| \leq 0.75$.

\subsection{Jet Production in Electron Positron Collisions}

In this section, we highlight renormalization-scale- and merging-scheme 
uncertainties by referring to electron-positron annihilation into jets.
Electron positron collisions are simulated at the center of mass energy 
$\sqrt{s} = M_\mathrm{Z}$, to be able to compare the different merging 
prescriptions to LEP data. We use a $p_{\perp,ij}^\mathrm{Lund}$ merging 
scale of value 5 GeV. 

\subsubsection{Comparison of UNLOPS-1, UNLOPS-P and UNLOPS-PC}

\begin{figure}
\centering
\includegraphics[width=0.5\textwidth]{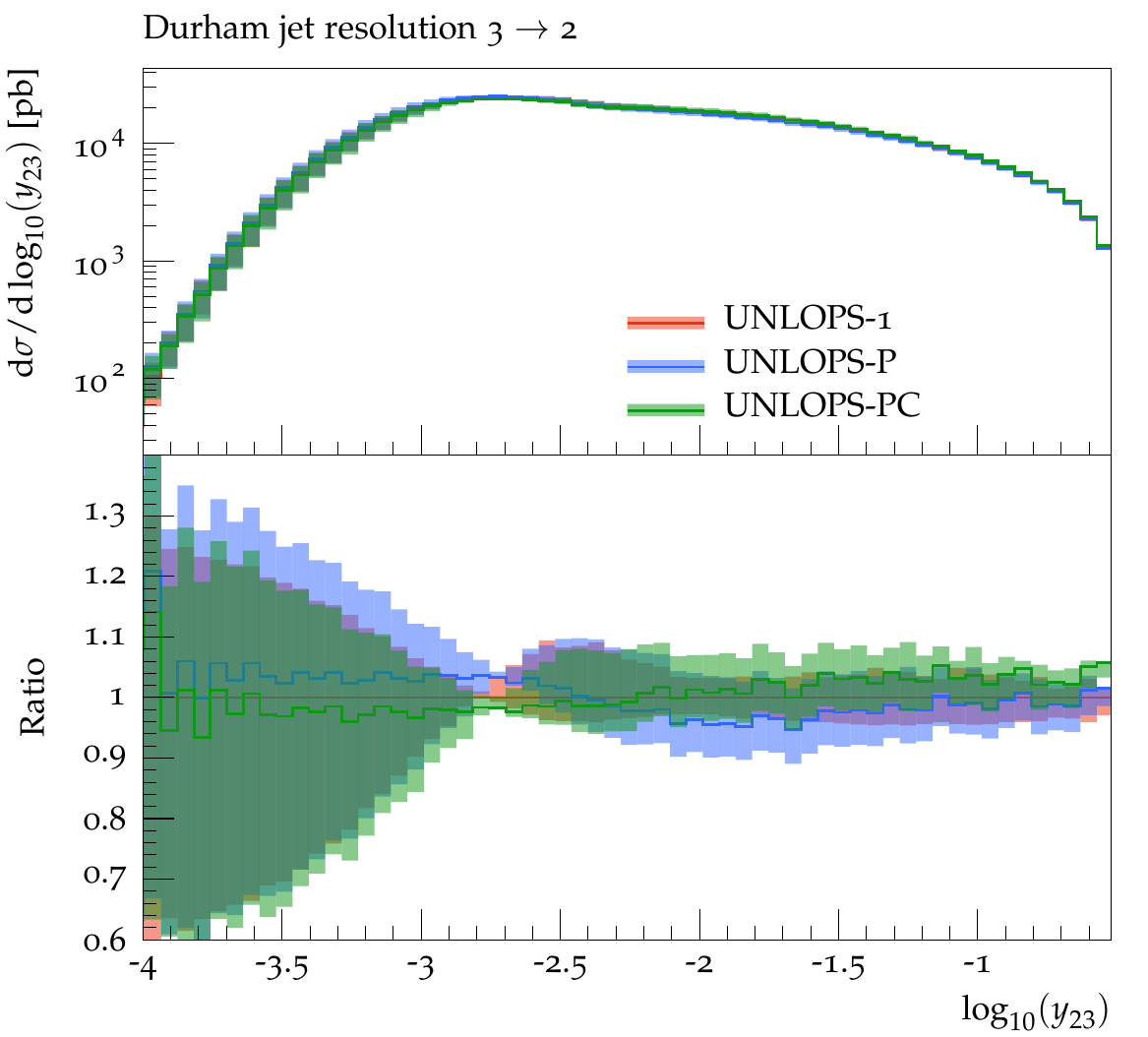}%
\includegraphics[width=0.5\textwidth]{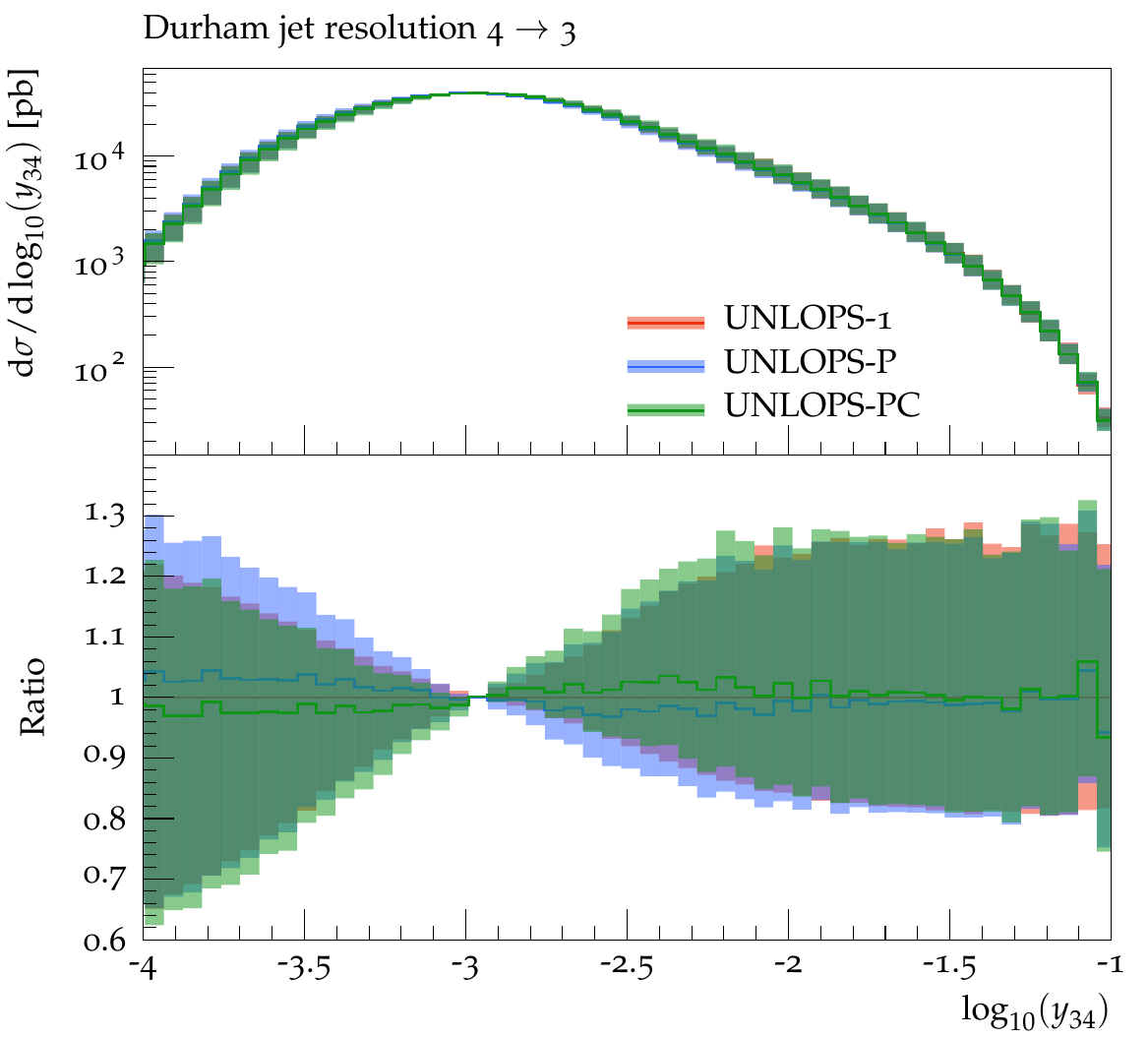}
\caption{Differential jet separation distributions $y_{23}$ and $y_{34}$ in $e^+ e^- \rightarrow \text{jets}$ at $\sqrt s = M_\mathrm Z$. The left distribution is described at NLO accuracy at high scales, the right at LO accuracy for all jet separations.}\label{fig:mergecomp}
\end{figure}

\Cref{fig:mergecomp} shows the differential jet separation distributions in the $3\rightarrow2$ clustering $y_{23}$ and in the $4\rightarrow3$ clustering $y_{34}$, with the Durham $k_\perp$-jet separation \cite{Catani:1991hj} 
\begin{equation}
  y_{ij} = 2(1-\cos \theta_{ij})\min(E_i^2,E_j^2)/s\, ,
\end{equation}
normalized with the squared CM energy $s$. The differing weighting 
prescriptions in the different schemes affect both the central prediction 
and the renormalization scale variation bands. 

The $y_{23}$ jet separation distribution in \cref{fig:mergecomp} shows
that the central prediction at NLO accuracy agrees between UNLOPS-1 and 
UNLOPS-P at high jet separations, since the Sudakov factor is close to unity
for in these regions, and no strong coupling ratios are applied in the two
schemes. In the UNLOPS-PC prescription, the strong coupling ratio introduces an 
upward shift, by effectively evaluating the coupling at a lower scale. Going to 
lower scales, UNLOPS-P falls compared to UNLOPS-1, due to the Sudakov 
suppression. In UNLOPS-PC, the strong coupling running counteracts this effect,
leading to a milder decrease. The unitary property in all schemes ensures that 
an increase (decrease) at high scales induces a decrease (increase) and lower 
scales. This leads to the observed central predictions at low separations 
behaving opposite to the high scale results for every scheme. High $y_{34}$ 
values agree for all schemes up, to statistical fluctuations, since in this
region, all schemes recover the result of the UMEPS unitary merging 
prescription \cite{Lonnblad:2012ng}. However, we observe differences at 
lower scales. This can be explained by the NLO precise lower multiplicity 
sample (here, the $e^+ e^- \rightarrow 3 \;\text{jet}$ NLO sample), that is 
modified by the different weighting prescriptions, and is showered below
the merging scale, thus contributing to $y_{34}$ at small separations. Overall, 
the central description of jet separation observables, which are very 
sensitive to the merging weight prescriptions, differs by up to 
about 5\% between the described schemes.

Scale variation bands for each merging scheme include variations of
fixed-order, parton-shower and merging-reweighting origin. As opposed to 
renormalization scale variations in the matrix elements only, this induces 
larger uncertainties at small jet separations, where emissions are generated 
by the parton shower. For observables at LO precision, e.g. $y_{34}$ in 
\cref{fig:mergecomp}, variations of the scales induce a very large band,
 amounting to about 20\% in each direction for all schemes alike. In unitary 
merging schemes, the subtraction of the respective jet multiplicity sample 
from the next-lower jet topology, turn the variation bands around, since they
contribute, via showering, at low jet separations. This leads to a region
with unphysically small uncertainty bands where the varied distributions cross.

Predominantly NLO precise distributions, such as $y_{23}$ in \cref{fig:mergecomp}, 
show smaller bands at high jet separations. In this region, renormalization 
scale uncertainties in the method contribute only at 
$\mathcal{O}(\alpha_\mathrm{s}^2)$ due to NLO-precise inputs, instead of 
$\mathcal{O}(\alpha_\mathrm{s})$ otherwise. At small jet separations, the 
distribution is again described by parton shower emissions instead, so that a 
large variation is observed. At the transition, there is an unphysically 
small variations, as observed for LO observables.

Comparing the size of the variation bands between UNLOPS-1, UNLOPS-P and 
UNLOPS-PC, we note that UNLOPS-1 and UNLOPS-PC are very similar, while 
UNLOPS-P leads to larger scale variation bands. This suggests that the 
application of Sudakov suppressions alone -- without taking strong coupling 
ratios into account -- introduces an additional variation, which is partly 
cancelled by the coupling ratio variation in UNLOPS-PC. The Sudakov variation 
behaves as $1-\alpha_\mathrm{s}(c_1 L^2 + c_2 L + c_3) + \mathcal{O}(\alpha_\mathrm{s}^2)$, 
while the $\alpha_\mathrm{s}$ ratio behaves as $1+\alpha_\mathrm{s} c^\prime L + \mathcal{O}(\alpha_\mathrm{s}^2)$,
where $L$ denotes a logarithmic enhancement of type $\ln{Q^2/p_\perp^2}$, $Q$ denotes 
a characteristic scale of the hard process and $p_\perp^2$ the scale of jet 
separation. At $\mathcal{O}(\alpha_\mathrm{s}^3)$ (i.e.\ induced by
$\mathcal{O}(\alpha_\mathrm{s})$ terms of the reweighting) we thus observe a partial 
cancellation in the single logarithmic contribution in UNLOPS-PC, which is not
present in UNLOPS-P. We have found changes of similar size when setting 
$w_\text{S}=1,~w_\text{NLO}=w_\text{LO}$, i.e.\ removing the consistency 
condition in eq.~\ref{eq:b1nlo_rwgt} and jeopardizing the logarithmic
accuracy. Thus, a very conservative uncertainty estimate that also 
acknowledges the potential of subleading-logarithmic mismodeling, should 
likely consist of the combination of the scale uncertainties of all three
merging schemes.

\subsubsection{Comparison to LEP Measurements}

\begin{figure}[h]
\centering
\includegraphics[width=0.5\textwidth]{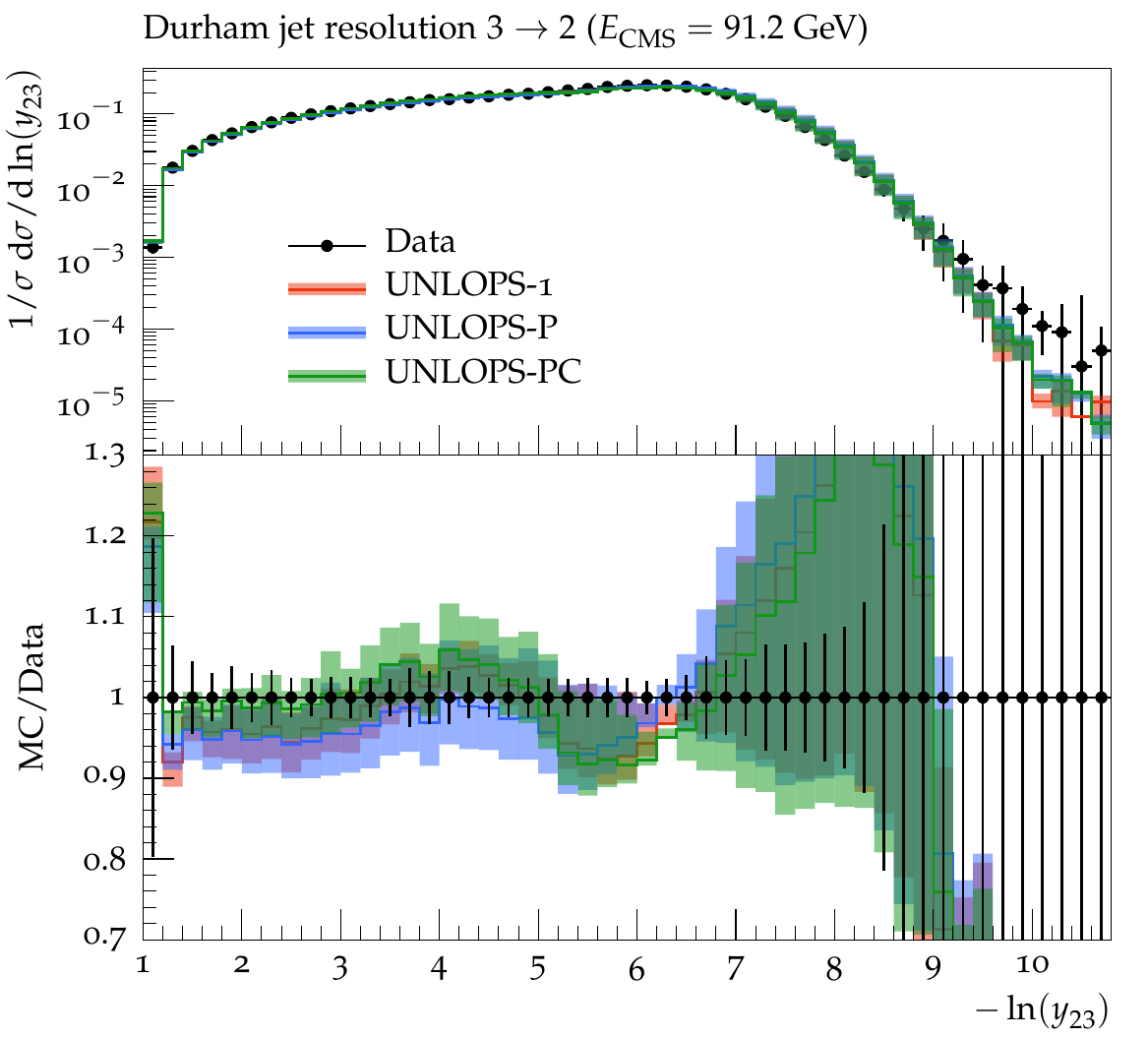}%
\includegraphics[width=0.5\textwidth]{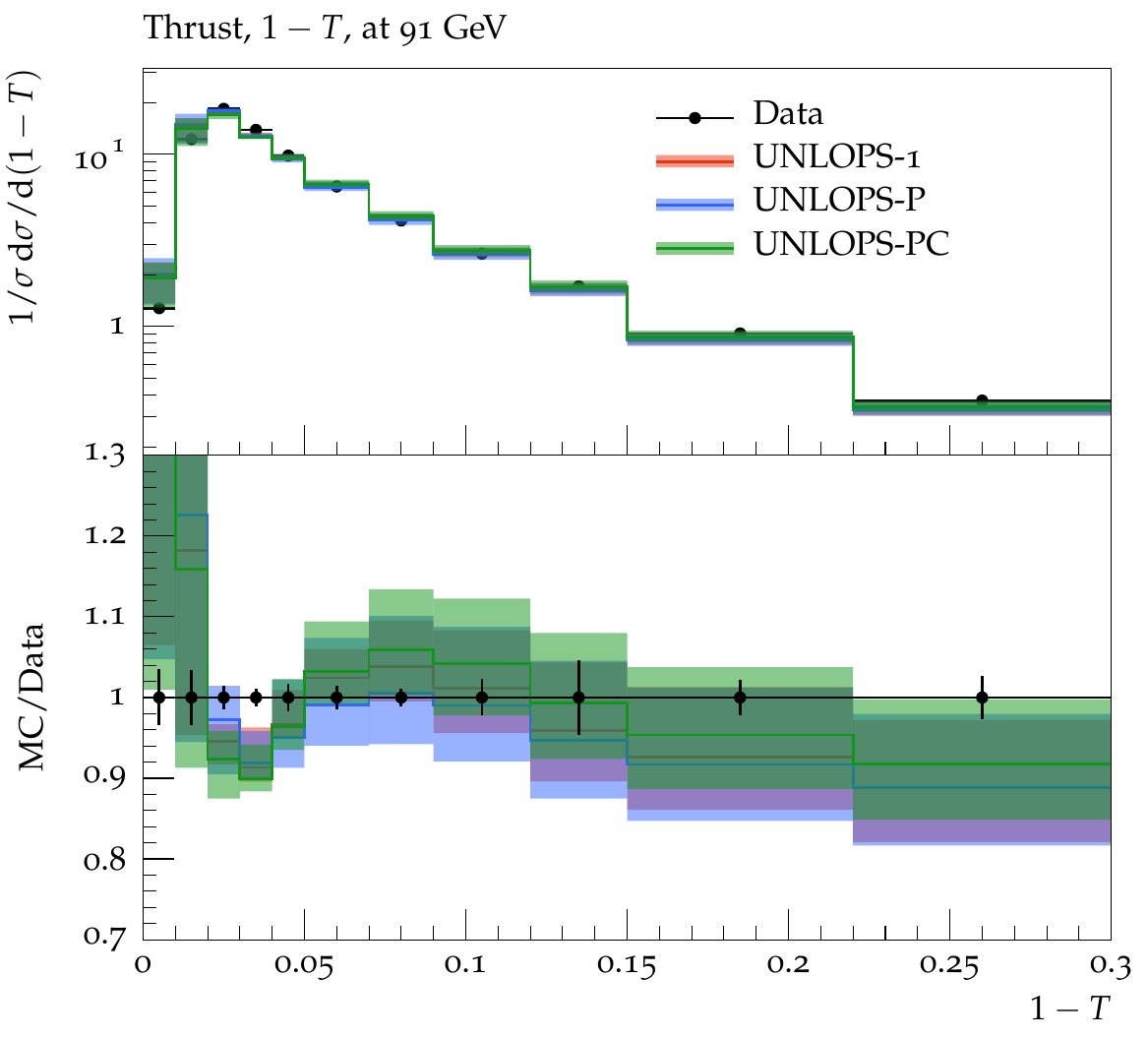}
\caption{Renormalization scale variation bands for the differential Durham jet resolution and thrust compared for the different variations of unitary NLO multi-jet merging. Data from \cite{Heister:2003aj} and \cite{Abbiendi:2004qz}.}\label{fig:eedata}
\end{figure}

As observed above, the difference between the weighting prescriptions is 
rather minor, amounting to no more than about 5\% for very sensitive jet 
separation observables. We furthermore do not observe a large difference in 
the description of experimentally measured data distributions. 
In \cref{fig:eedata}, we compare the thrust and Durham jet resolution to 
ALEPH \cite{Heister:2003aj} and OPAL \cite{Abbiendi:2004qz} data.  

The Durham jet separation distribution is well described by all schemes, 
especially at large jet separations. In total, the data is compatible with the 
prediction of all schemes across most jet separations. Only at very low 
separations, where the statistical uncertainty on the data is rather large, 
do the prediction differ mildly from the data distribution. Overall, 
UNLOPS-PC agrees slightly better with the measured data distribution, compared 
to the other weighting schemes.

The thrust distribution
\begin{equation}
  1-T = 1 - \max_{\vec n} \frac{\sum_i |\vec n\cdot \vec p_i|}{\sum_i |\vec p_i|}
\end{equation}
ranges between zero for very narrow back-to-back jet configurations, 
corresponding to very soft ``hardest'' emissions, and $1/3$, corresponding to 
three very well separated jets. We find larger scale variations -- caused by 
shower emissions -- at low $1-T$, while high $1-T$ variations are milder. The 
band is in general wider than in the Durham jet separation distribution, since 
the thrust distribution is more sensitive to further emissions described only 
at LO. The agreement with data is satisfactory, with differences only 
at low $1-T$, where even the ``hardest'' emission is modeled
solely by the parton shower.

In both cases, the unphysically narrow variation bands where the variations cross
can potentially lead to a significant deviation from measured data. In the 
observables shown in \cref{fig:eedata}, an envelope of the result of all 
schemes may be used to mitigate such an effect. 

\subsection{W+jets Production in Proton Proton Collisions at LHC}
\label{sec:wjets}

\begin{figure}
\centering
\includegraphics[width=0.5\textwidth]{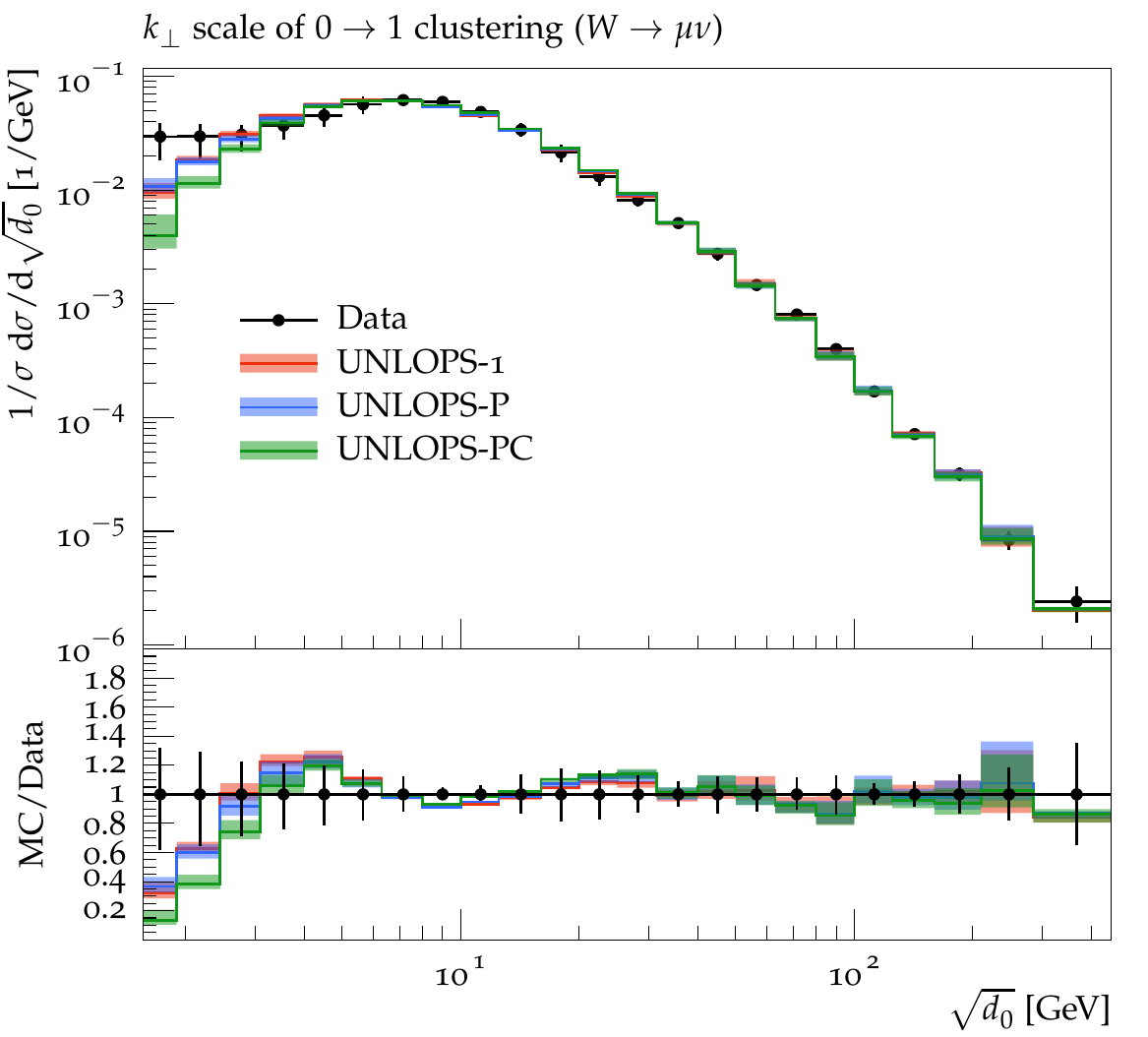}%
\includegraphics[width=0.5\textwidth]{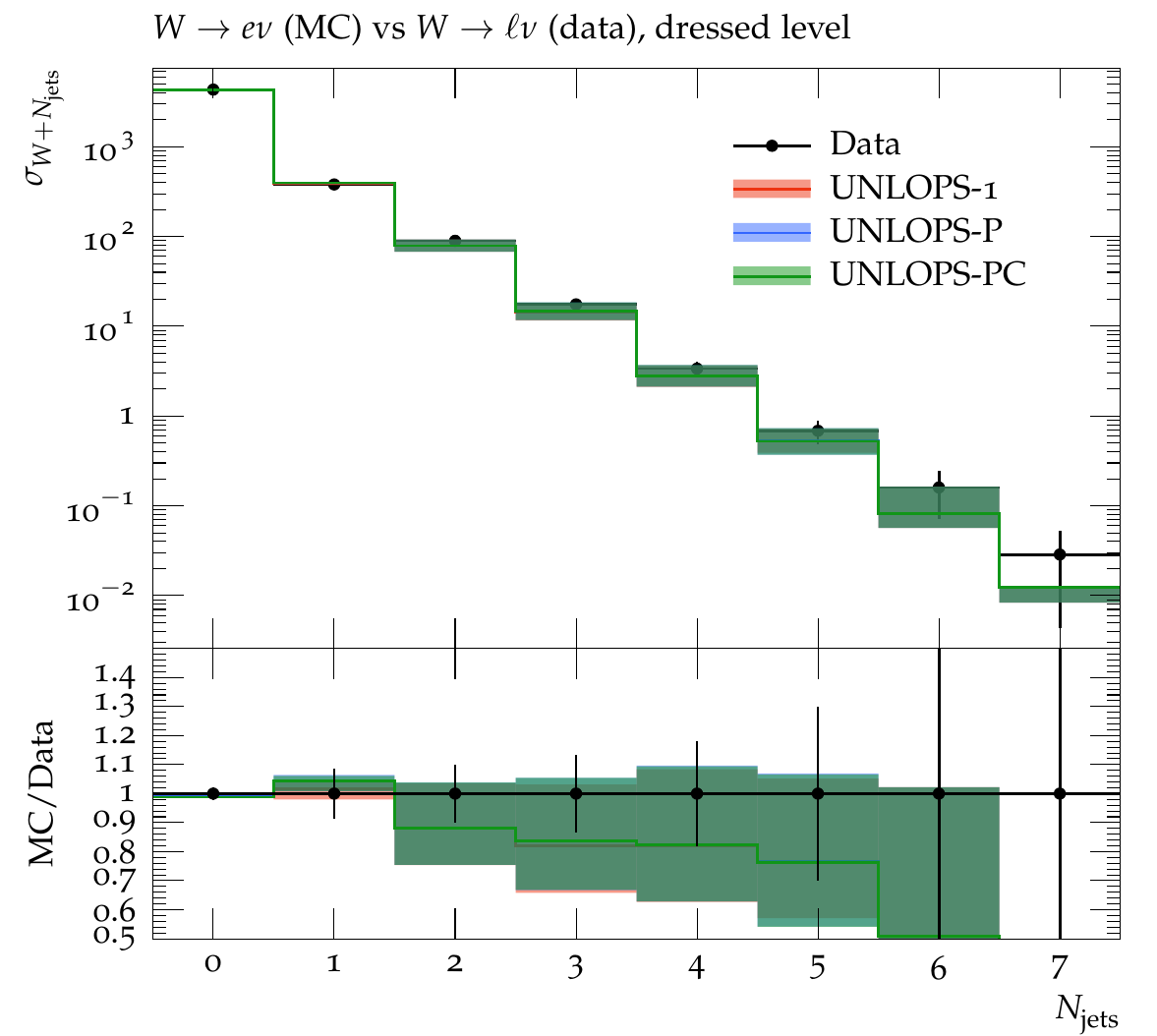}
\caption{W + jets production at proton proton collisions with $\sqrt s = 7$ TeV: $k_\perp$ splitting scale of first jet \cite{Aad:2013ueu} and exclusive jet multiplicity \cite{Aad:2014qxa}.}\label{fig:wjetcomp}
\end{figure}

The simulation of final states in hadron collisions is typically 
much more involved than in lepton collisions, due to the rich structure of
the composite colliding particles, as well as the larger phase-space available
to the final-state particles. Thus, an assessment of uncertainties due to
scale- and scheme variations in hadron colliders is necessary.
In this section, we illustrate these uncertainties using $W$ + jets final states
at proton-proton collisions at $\sqrt{s} = 7 \,\text{TeV}$. Jet observables of 
this process are then compared to LHC data. We use a $p_{\perp,ij}^\mathrm{Lund}$ 
merging scale definition, with value $10 \,\text{GeV}$ as merging scale 
cut. Furthermore, we apply a NLO $K$-factor 
$K = \sigma_\mathrm{NLO}(pp\rightarrow W)/\sigma_\mathrm{LO}(pp\rightarrow W)$ 
to all leading-order input configurations that could have otherwise been reached by the 
$p_\perp$ ordered shower. Non-ordered additional jet configurations are thus 
interpreted as ``genuine" real-emission corrections, for which a naive
rate correction can be considered questionable. All results are produced using the 
NNPDF3.1\_nlo\_as\_0118 PDF set \cite{Ball:2017nwa} via the LHAPDF 
framework \cite{Buckley:2014ana}.

Proton-proton collisions introduce (at least) two more sources of
renormalization scale uncertainty: The treatment of running couplings
in initial-state shower evolution, and in multiparton interactions.
Since the latter are highly correlated with other semi- or
non-perturbative parameters, we do not consider their impact in this
study. Renormalization-scale and merging-scheme variations for hadronic
initial states thus require only simple generalizations on top of the 
previous section. 

Since the term in parentheses in 
eqs.~\ref{eq:unlops1}$-$\ref{eq:unlopspc} acquires PDF-dependent
components, it is not obvious that the level of similarity found 
in $e^+e^-$ collisions is also present at hadron colliders.
\Cref{fig:wjetcomp} shows the weighting schemes compared to 
$pp\rightarrow W+\text{jets}$ data at $\sqrt{s}=7\,\text{TeV}$ for the 
$k_\perp$ $0\rightarrow 1$ clustering scale \cite{Aad:2013ueu} and the 
exclusive jet multiplicity \cite{Aad:2014qxa}.
At high $\sqrt{d_0}$, no difference between the schemes is found. This is due to
the reference renormalization scale $\mu_\mathrm{R} = M_\mathrm{W}$ chosen in 
the generation of matrix elements being reachable by parton showering. Thus,
the differences between  UNLOPS-P and UNLOPS-PC are less pronounced than 
in $e^+e^-$ collisions. We observe a very light suppression of UNLOPS-PC at high 
$\sqrt{d_0}$, compared to UNLOPS-P. The strong coupling ratio enhances 
UNLOPS-PC compared to UNLOPS-P below $M_\mathrm{W}$. At very low scales, the 
distribution is again dominated by parton shower emissions from 
$0$-parton states, as well as shower emissions from the integrated subtraction 
of the NLO 1-jet sample. 
The overall strong coupling ratio enhancement of the subtraction in UNLOPS-PC is
consistent with the lower  UNLOPS-PC result observed at low scales. Note
that with the chosen PDF set, all schemes struggle to describe the 
low-$p_\perp$ region satisfactorily. This suggests that a retuning of the 
MPI model might be necessary when using this setup productively.

The first jet clustering scale $\sqrt{d_0}$ is dominated by NLO-precise
contributions at high values, and thus has a very small scale variation band
in that region. However, this seems to also be the case at small separation,
where the parton shower, as well as MPI effects, dominate. The reason for these
milder variations lies in the implementation of the shower scale variations in 
\pythia 8 \cite{Mrenna:2016sih}: In order to avoid numerical instabilities in the reweighting procedure when 
approaching low scales, no shower variations are performed below a 
certain scale (determined by multiplication of a factor and the shower 
cut-off scale). In ISR, this is applied to the regularization 
parameter {\tt pT0Ref} with default value of $2$ GeV, while in FSR,
it applies to the {\tt pT0} parameter of default value $0.5$ GeV. Thus, shower 
variations in ISR are suppressed below transverse momentum scales of about 6 
GeV. These values were chosen to limit the size of the weight fluctuations 
induced by the shower reweighting procedure of~\cite{Mrenna:2016sih}, but 
lead to the merging prescriptions not agreeing within their bands at low
scales. Using different merging schemes thus helps to isolate phase-space
regions with questionable uncertainty estimates, and a combination of 
scheme variations is advisable.

The right plot in \cref{fig:wjetcomp} shows the exclusive jet multiplicity of 
jets with $p_\perp > 30\,\text{GeV}$ \cite{Aad:2014qxa}. The 0-jet 
and 1-jet bins, which are described with NLO precision, are reproduced very 
well. Higher multiplicities, described at LO or parton-shower accuracy only, 
are underestimated. While the differences between UNLOPS-P and UNLOPS-PC 
are negligible in this observable, both schemes yield a very slightly 
larger exclusive 1-jet rate than UNLOPS-1. If the contribution of the term in 
square brackets in \cref{eq:unlops1,eq:unlopsp,eq:unlopspc} was mostly 
positive in the relevant region of phase space, the Sudakov factor should 
rather lead to a lower prediction for UNLOPS-PC and UNLOPS-P. That this is
not the case suggest that the contribution is negative at least in some 
parts of the phase space, highlighting that there is a non-trivial
interplay between the different contributions, and that rule-of-thumb reasoning
should be considered with caution.

\begin{figure}
\centering
\includegraphics[width=0.33\textwidth]{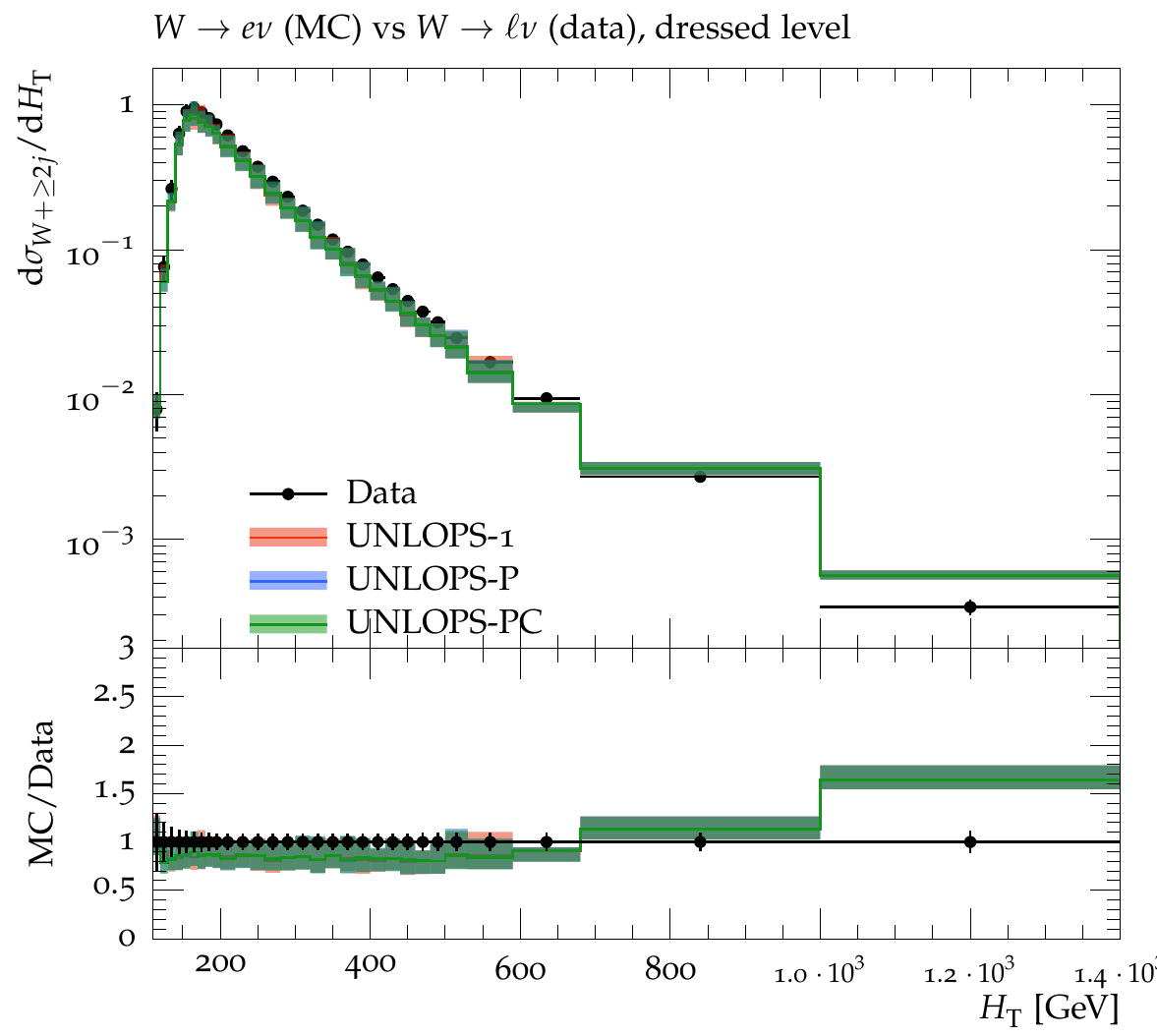}%
\includegraphics[width=0.33\textwidth]{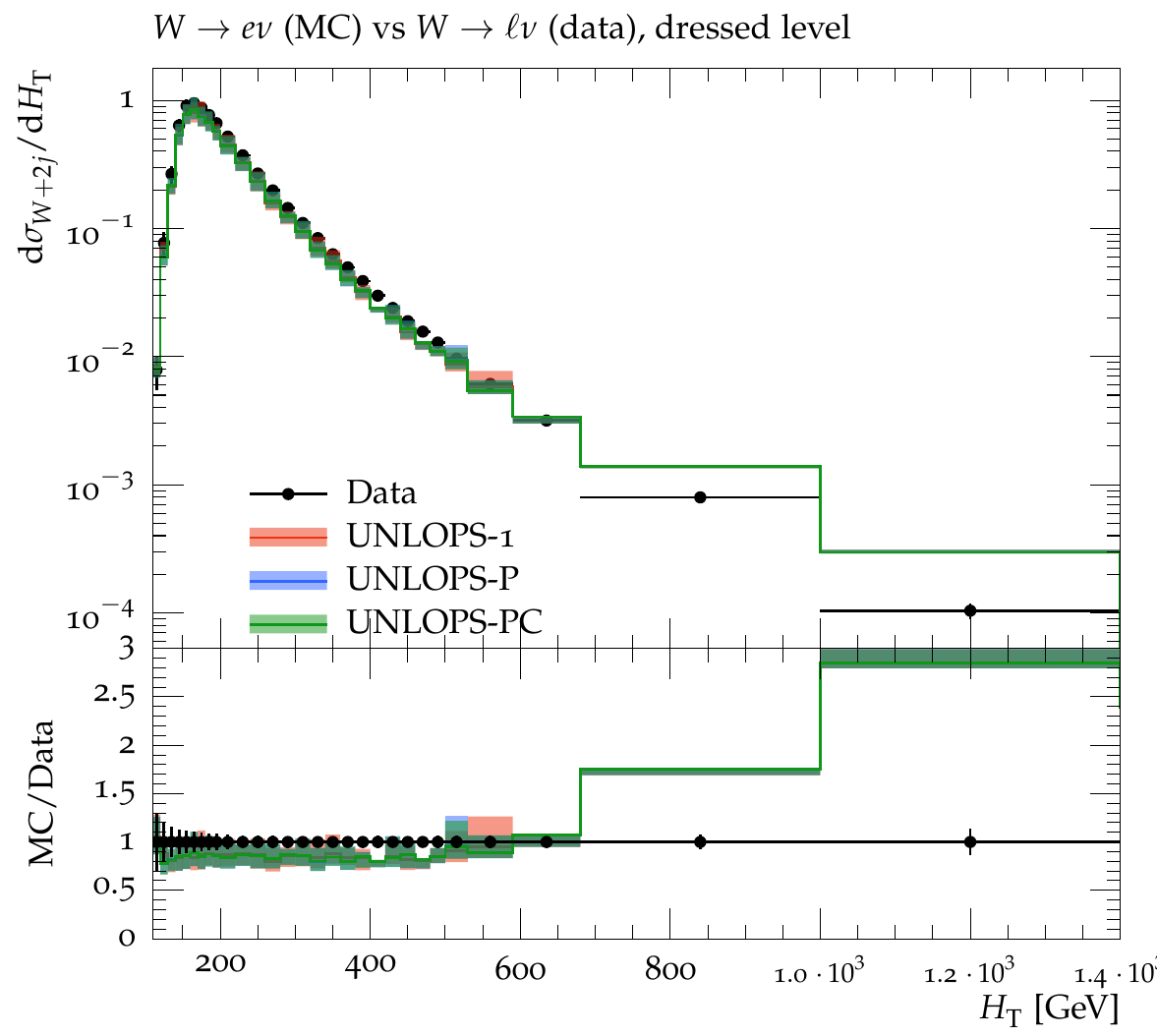}%
\includegraphics[width=0.33\textwidth]{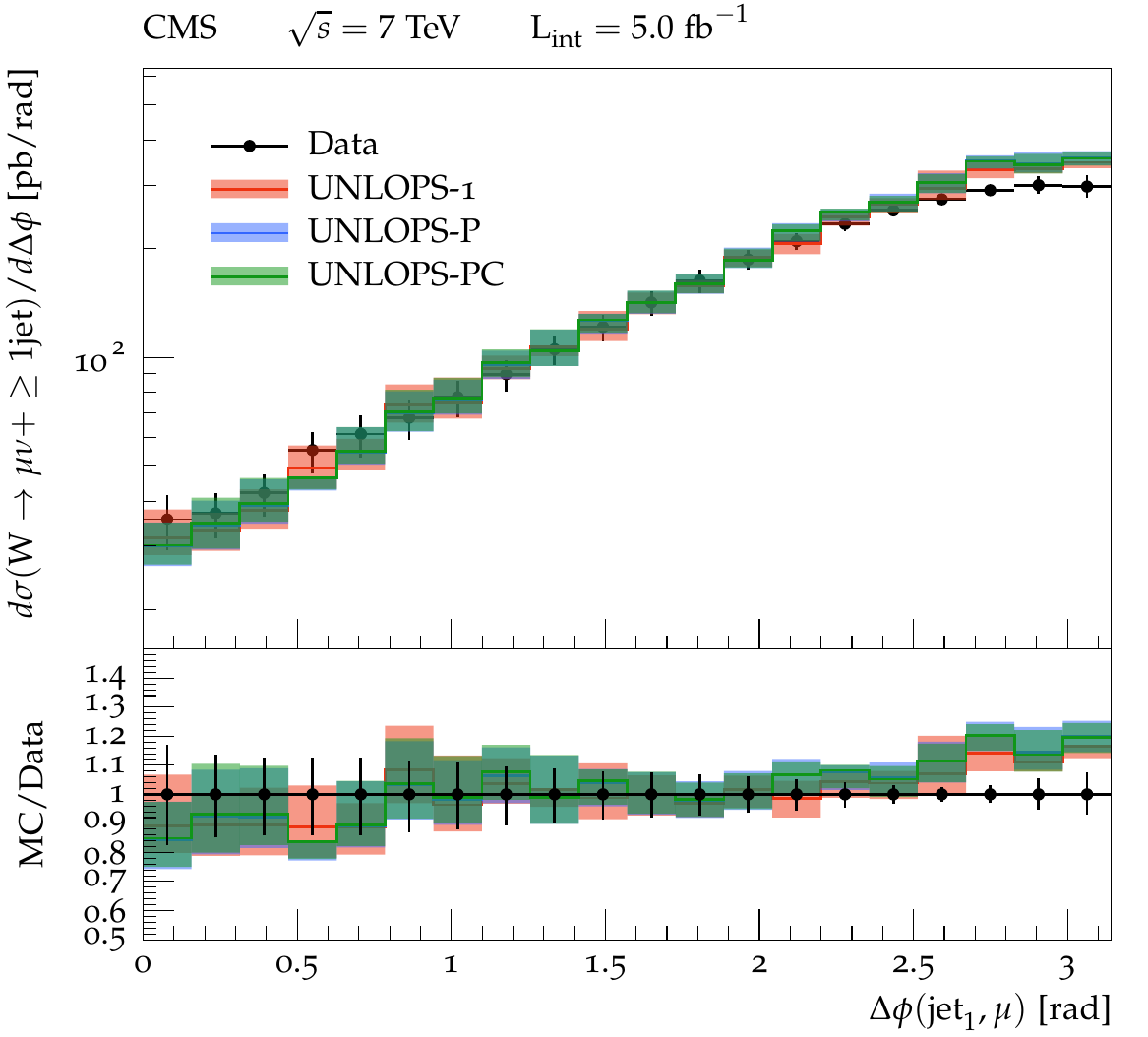}
\caption{W + jets production at proton proton collisions with $\sqrt s = 7$ TeV: Scalar sum of jet transverse momenta $H_\mathrm{T}$ for inclusive and exclusive 2 jet events \cite{Aad:2014qxa} and azimuthal distance $\Delta\phi$ between hardest jet and muon \cite{Khachatryan:2014uva}.}\label{fig:wjethtphi} 
\end{figure}

The left two plots in \cref{fig:wjethtphi} show the scalar sum of jet 
transverse momenta $H_\mathrm{T}$ for inclusive and exclusive 2-jet events. 
In particular distribution of $H_\mathrm{T}$ in exclusive 2-jet events shows a strong 
overshooting of the prediction, compared to the data. This is due
to a mismodeling of the prediction for the transverse momenta of the first- and 
second-hardest jets. In inclusive 2 jet events, this effect is milder due to
an underestimate of subleading jet transverse momenta, which conspires
with the former effect to yield a less pronounced effect. Appropriate scale 
choices for unordered jet event topologies~\cite{Fischer:2017yja} or 
the inclusion of electro-weak histories~\cite{Schalicke:2005nv,Christiansen:2015jpa}
have been inferred to improve this situation\footnote{We use the \pythia settings Merging:unorderedASscalePrescrip=1 to use a combined scale setting in $\alpha_\mathrm{s}$ for unordered histories and Merging:IncompleteScalePrescrip=1 to get sensible shower starting scales for incomplete histories.}. The new NLO merging prescriptions 
proposed in this study do not improve this mismodeling of data. Finally, 
the right plot in \cref{fig:wjethtphi} shows the azimuthal distance between 
the hardest jet and the muon in leptonic $W$ signatures. No significant 
differences between the schemes is observed, suggesting that
the correlation between QCD- and electroweak parts of the events are insensitive
to the scheme variation.

The observables discussed here serve as an illustrative example of the effects of scale and scheme variations in unitary NLO merging. Similar effects can be seen in other observables. More observables, processes and energies can be studied with the implementation made available in a future release of \pythia 8.3.

\section{Summary and Outlook}
\label{sec:outlook}

Event generator uncertainties are one of the main obstacles to precision
measurements in collider experiments. This is particularly obvious when
event generators are used as an accurate model of large backgrounds to
high-energy signal processes with low rate appearing in the tails of
Standard-Model-dominated observables. To describe such backgrounds, precise
NLO merged event generator calculations are needed. Variations of perturbative
parameters in such calculations should give a good indication of the overall
precision of the predictions. However, since NLO merged calculations
include fixed-order as well as all-order effects, and combine multiple 
calculations in an intricate manner, it is not completely obvious how a 
realistic perturbative uncertainty should be assessed. In this article, we
have presented steps towards this goal, and in particular focused on the
interplay between renormalization scale choices and the very definition of
the merging scheme at higher orders. These enter at the same order, such 
that it is important to quantify their individual impact, as well as their
correlation. For this purpose, we extended the unitarized NLO merging
prescription in \pythia 8 to accommodate renormalization scale 
variations (as a combined framework encompassing correlated 
variations of fixed-order, parton-shower and merging components) and
merging scheme changes. For the latter, we have introduced two extensions
of the UNLOPS method, which were motivated by different interpretations
of dressing process-dependent NLO corrections with the all-order effects
of soft and collinear radiation. The implementation will be publicly available
in a future release of the \pythia 8.3 event generator.

The renormalization-scale and merging-scheme variations were used to estimate
the perturbative uncertainty of illustrative observables in electron-positron and
proton-proton collisions. Overall, the estimate is as expected: The uncertainty
is small in regions that are primarily sensitive to the
NLO fixed-order components of the of the calculation, and large in regions 
dominated by parton showering. The renormalization scale uncertainty bands of
different merging schemes largely overlap. In transition regions
between calculations of different jet multiplicity, the difference between
(the central result of) the schemes can be larger than scale variations, due
to the latter being artificially small due to unitarity requirements.
Some visible differences between the merging schemes, in size similar to scale 
uncertainties, can also remain in regions sensitive to very well-separated 
jets. This is mainly related to a different definition of the functional form
of the argument of the running coupling at higher orders, which persists even
in those phase space regions. Thus, a joint scale- and
scheme- variation may be considered a more reliable uncertainty
estimate. 

The current study should be regarded as an initial step in the assessment
of uncertainties in unitarized NLO merging schemes. We have focused on 
a subset of variations for which a minimization of contamination by 
statistical fluctuations in the event generation is possible through
a reweighting procedure. It would be very valuable to extend this property also
to other sources of uncertainty, such as consistent combined 
factorization-scale and shower starting-scale variations, changes in 
cut-off of parton-shower evolution, and changes in the merging scale
definition and value. These would require an extensive redesign of the 
parton-shower algorithm. More insight into the effect of using different PDF 
parametrizations would also be valuable, but is complicated by the strong
correlation with other semi- or non-perturbative components of the 
event generator -- which is commonly held fixed after event generator tuning.
Nevertheless, we believe that making the developments presented in the
current study available within the \pythia 8 event generator will already
allow also non-developers to perform more 
systematic studies of Standard-Model background uncertainties.

\section{Acknowledgments}

We would like to thank Peter Skands for discussions and encouragement at an 
early stage in the project, and Malin Sj{\"o}dahl for collaboration at an 
early stage, and for continuous feedback. 
We also thank Ioannis Tsinikos for careful reading and useful comments on the manuscript.
This work has received funding from the European Union's Horizon 2020 research and innovation programme as part of the Marie Sk{\l}odowska-Curie Innovative Training Network MCnetITN3 (grant agreement no. 722104).

\appendix 

\section{Implementation Details}

\subsection{Merging Weight Variation}

For completeness, we describe some details of the implementation of scale variations, especially for the $\mathcal{O}(\alpha_\mathrm{s})$ contribution.

We generate the matrix element samples with a fixed renormalization scale. At leading order, we do not use scale variations in the matrix element samples, since $\alpha_\mathrm{s}$ variations can simply be implemented as strong coupling ratios in \pythia based on the jet multiplicity of the event.

The all order weights are calculated as described above. For leading order samples, the strong coupling ratio has the central scale in the denominator, to be consistent with no variations in the leading order matrix element samples. Emission probability variations are generated from weight variations in the parton shower. The PDF ratio, MPI weights and the K factor are not varied.

The expanded weights to $\mathcal{O}(\alpha_\mathrm{s})$, which are only applied to leading order matrix element input, can be written as
\begin{equation}
  B_n(\mu_\mathrm{R})\left(\frac{\alpha_\mathrm{s}(k\mu_\mathrm{R})}{\alpha_\mathrm{s}(\mu_\mathrm{R})}\right)^n\left(1 + \frac{\alpha_\mathrm{s}(k\mu_\mathrm{R})}{\alpha_\mathrm{s}(\mu_\mathrm{R})}(\text{First term in $\alpha_\mathrm{s}(\mu_\mathrm{R})$ expansion of weight factors}) + \{K\}_{\alpha_\mathrm{s}(\mu_\mathrm{R})}\right)\, ,
\end{equation}
with jet multiplicity $n$. The first order contribution to the strong coupling ratio is $\alpha_\mathrm{s}(k\mu_\mathrm{R}) b_0/(2\pi)0.5\log(\mu_\mathrm{R}^2/p_\perp^2)$. The variation in the logarithm cancels since it is applied to both the renormalization scale and the shower scale. For other weight components, only the $\alpha_\mathrm{s}$ coefficient is varied. The $K$ factor is not varied.

\subsection{$e^+ e^-$ Jet Cut}

The $p_\perp^\mathrm{Lund}$ cut we employ as merging scale cut is not available for NLO matrix elements in MG5\_aMC. Instead, a sufficiently inclusive jet $k_\perp$ cut can be used to regularize collinear divergences, and the $p_\perp^\mathrm{Lund}$ cut is then applied in \pythia 8. To make sure that this alternative $k_\perp$ cut is not stronger than $p_\perp^\mathrm{Lund}$ for specific configurations, the $k_\perp$ value is usually chosen much smaller than the desired merging cut, leading to a lower efficiency.

For electron positron collisions, a Durham $k_\perp$ jet cut \cite{Catani:1991hj}, denoted as $d_{ij}$, can be used to regularize the $+1\,\text{jet}$ NLO matrix elements with the same value as the Lund $p_\perp$ \cite{Sjostrand:2004ef} merging scale. The requirement for this to be allowed is $d_{ij} \geq p_{\perp,ij}^\mathrm{Lund}$ such that a cut on $d_{ij}$ is more inclusive than a $p_\perp^\mathrm{Lund}$ cut. The Lund shower $p_\perp$ is given by
    \begin{equation}
      p_{\perp,ij}^\mathrm{Lund} = z(1-z) q^2 = \frac{E_i E_j}{(E_i + E_j)^2} (m_i^2 + m_j^2 + 2 (E_i E_j -|\vec p_i| |\vec p_j| \cos\theta_{ij})) \, .
    \end{equation}
    Here we use the angle $\theta_{ij}$ between partons $i$ and $j$ and the energy fractions $z$ and $1-z$ as employed by the \pythia $p_\perp$ shower. If we generate events with zero quark masses, we find
    \begin{align}
      p_{\perp, ij}^\mathrm{Lund} &= \frac{2 E_i^2 E_j^2}{(E_i + E_j)^2} (1-\cos\theta_{ij}) = \frac{2\min(E_i^2,E_j^2)\max(E_i^2,E_j^2)}{(\min(E_i,E_j) + \max(E_i,E_j))^2}(1-\cos\theta_{ij}) \\
      &\leq \frac{2\min(E_i^2,E_j^2)\max(E_i^2,E_j^2)}{(\max(E_i,E_j))^2}(1-\cos\theta_{ij}) = 2\min(E_i^2, E_j^2)(1-\cos\theta_{ij}) = d_{ij} \, ,
    \end{align}
which justifies the efficient jet $d_{ij}$ cut on the generated $+1\,\text{jet}$ NLO matrix element samples. However, the above is only true for the first emission: The energy ratios in the Lund $p_\perp$ measure are defined in the dipole center of momentum frame, while the energies in the Durham clustering are taken in the whole event center of momentum frame. For the first emission, these two are identical. For further NLO jet corrections, which we do not employ here, and for proton proton collisions, a more conservative cut must be applied.

\bibliography{ref}{}
\bibliographystyle{h-physrev}

\end{document}